\documentclass[compsoc,conference,a4paper,10pt,times]{IEEEtran}

\usepackage[T1]{fontenc}
\usepackage[utf8]{inputenc}
\usepackage{microtype}

\usepackage{amsmath,amssymb,amsfonts,mathtools}
\usepackage{graphicx}
\usepackage{booktabs}
\usepackage{multirow}
\usepackage{siunitx}

\DeclareSIPrefix\tebi{Ti}{40}
\DeclareSIUnit\byte{B}

\usepackage{xspace}
\usepackage{balance}

\usepackage{enumitem} %

\usepackage{placeins} %

\usepackage{comment} %

\usepackage[
  backend=biber,
  style=ieee,
  natbib=true,       %
  citestyle=numeric-comp,  %
  maxbibnames=1,     %
  minbibnames=1,     %
  maxcitenames=2,    %
  mincitenames=1     %
]{biblatex}

\usepackage{csquotes}               %

\usepackage[hidelinks]{hyperref}
\usepackage{xurl}   %
\usepackage[capitalise,nameinlink]{cleveref}

\usepackage[main=english, provide+=*]{babel}
\babelprovide[import]{latin}

\makeatletter
\addto\extraslatin{\itshape}
\makeatother

\newif\ifDRAFT
\DRAFTtrue

\usepackage{xcolor}
\ifDRAFT
  \newcommand{\todo}[1]{\textcolor{red}{\textbf{[TODO: #1]}}}
  \newcommand{\note}[1]{\textcolor{blue}{\textbf{[note: #1]}}}

\else
  \newcommand{\todo}[1]{}
  \newcommand{\note}[1]{}

\fi

\newcommand{\Mctrl}{\ensuremath{M_{\mathrm{ctrl}}}\xspace}      %

\newcommand{\effect}{\ensuremath{\tau}\xspace}
\newcommand{\effectd}{\ensuremath{\tau(d)}\xspace}

\newcommand{\doc}{\ensuremath{d}\xspace}

\newcommand{\auc}{\textsc{auc}\xspace}

\newcommand{\tprat}[1]{\ensuremath{\textsc{tpr}@#1\%\textsc{fpr}}\xspace}

\usepackage[
  acronym,
  nomain,
  nonumberlist,
  nohypertypes={main,acronym}, %
  notranslate,
]{glossaries-extra}

\setabbreviationstyle[acronym]{long-short}
\newacronym[longplural={Debugging Information Entities}]{DIE}{DIE}{Debugging Information Entity}
\newacronym[shortplural={OSes}, firstplural={operating systems (OSes)}]{OS}{OS}{operating system}

\newacronym{IQL}{IQL}{Independent Q‑Learning}

\newacronym{AI2}{AI2}{Allen Institute for AI}
\newacronym{API}{API}{application programming interface}
\newacronym{AUC}{AUC}{area under the ROC curve}
\newacronym{CI}{CI}{confidence interval}
\newacronym{DPO}{DPO}{direct preference optimization}
\newacronym{FDR}{FDR}{false-discovery rate}
\newacronym{FPR}{FPR}{false-positive rate}
\newacronym{LLM}{LLM}{large language model}
\newacronym{MI}{MI}{membership-inference}
\newacronym{RMIA}{RMIA}{Robust Membership Inference Attack}
\newacronym{ROC}{ROC}{receiver operating characteristic}
\newacronym{RLHF}{RLHF}{reinforcement learning from human feedback}
\newacronym{RLVR}{RLVR}{reinforcement learning with verifiable rewards}
\newacronym[
  shortplural={SDs},
  longplural={standard deviations}
]{sd}{SD}{standard deviation}
\newacronym{SFT}{SFT}{supervised fine-tuning}
\newacronym{TPR}{TPR}{true-positive rate}

\makeatletter
\renewcommand\paragraph{%
  \@startsection{paragraph}{4}%
  {0pt}%
  {0ex \@plus 0.1ex \@minus 0.1ex}%
  {-1em}%
  {\normalfont\normalsize\bfseries}%
}
\makeatother

\newif\ifANON
\ANONfalse
\begin{document}
\pagestyle{plain}  %

\title{From Training Exposure to Membership Evidence:\\A Controlled Study of OLMo~2}

\ifANON
  \author{\IEEEauthorblockN{Anonymous Submission}}
\else
  \author{
    \IEEEauthorblockN{
      Ali Banaei,
      Muhammad Shahab Shahzad,
      Gerald Q. Maguire Jr.,
      Marco Chiesa,
      Dejan Kosti\'c
    }
    \IEEEauthorblockA{
      KTH Royal Institute of Technology\\
      Stockholm, Sweden\\
      \{alibma, mssh, maguire, mchiesa, dmk\}@kth.se
    }
  }

\fi

\maketitle
\thispagestyle{plain}

\newcommand{\npairsdesign}{7814}
\newcommand{\npairs}{7803}
\newcommand{\npairsnat}{4832}
\newcommand{\npairsE}{1994}
\newcommand{\taunat}{+0.0221}
\newcommand{\taunatlo}{+0.0212}
\newcommand{\taunathi}{+0.0229}
\newcommand{\senat}{0.0004}
\newcommand{\dznat}{0.71}
\newcommand{\sdnat}{0.0309}
\newcommand{\posnat}{85.4}
\newcommand{\aucnat}{0.767}
\newcommand{\aucnatlo}{0.759}
\newcommand{\aucnathi}{0.774}
\newcommand{\tpronenat}{3.8}
\newcommand{\tprpointonenat}{0.19}
\newcommand{\negnat}{4832}
\newcommand{\tauall}{+0.0335}
\newcommand{\taualllo}{+0.0319}
\newcommand{\tauallhi}{+0.0353}
\newcommand{\seall}{0.0009}
\newcommand{\dzall}{0.45}
\newcommand{\sdall}{0.0741}
\newcommand{\posall}{89.1}
\newcommand{\aucall}{0.793}
\newcommand{\aucalllo}{0.788}
\newcommand{\aucallhi}{0.799}
\newcommand{\tproneall}{7.3}
\newcommand{\tprpointoneall}{3.51}
\newcommand{\negall}{7326}
\newcommand{\tauefam}{+0.0226}
\newcommand{\tauefamlo}{+0.0216}
\newcommand{\tauefamhi}{+0.0235}
\newcommand{\seefam}{0.0005}
\newcommand{\dzefam}{1.03}
\newcommand{\sdefam}{0.0219}
\newcommand{\posefam}{96.6}
\newcommand{\aucefam}{0.957}
\newcommand{\aucefamlo}{0.951}
\newcommand{\aucefamhi}{0.963}
\newcommand{\tproneefam}{28.8}
\newcommand{\tprpointoneefam}{3.36}
\newcommand{\negefam}{1994}
\newcommand{\permz}{40}
\newcommand{\permp}{<10^{-5}}
\newcommand{\npower}{13}
\newcommand{\npowerall}{31}
\newcommand{\fdrfive}{0}
\newcommand{\fdrtwenty}{119}
\newcommand{\ppvprev}{0.8}
\newcommand{\collorcmone}{0.766}
\newcommand{\collorcmtwo}{0.797}
\newcommand{\collorcmthree}{0.836}
\newcommand{\collorcmfive}{0.884}
\newcommand{\collorcmonezero}{0.949}
\newcommand{\collorcmtwozero}{0.990}
\newcommand{\collorcmthreeone}{0.998}
\newcommand{\collorcmfivezero}{1.000}
\newcommand{\collorcmonezerozero}{1.000}
\newcommand{\collsibmone}{0.745}
\newcommand{\collsibmtwo}{0.787}
\newcommand{\collsibmthree}{0.820}
\newcommand{\collsibmfive}{0.871}
\newcommand{\collsibmonezero}{0.945}
\newcommand{\collsibmtwozero}{0.987}
\newcommand{\collsibmthreeone}{0.997}
\newcommand{\collsibmfivezero}{1.000}
\newcommand{\collsibmonezerozero}{1.000}
\newcommand{\collpubmone}{0.605}
\newcommand{\collpubmtwo}{0.644}
\newcommand{\collpubmthree}{0.669}
\newcommand{\collpubmfive}{0.706}
\newcommand{\collpubmonezero}{0.771}
\newcommand{\collpubmtwozero}{0.842}
\newcommand{\collpubmthreeone}{0.907}
\newcommand{\collpubmfivezero}{0.954}
\newcommand{\collpubmonezerozero}{0.985}
\newcommand{\tierpubtau}{+0.0220}
\newcommand{\tierpubdz}{0.27}
\newcommand{\tierpubauc}{0.603}
\newcommand{\tierpubtpr}{1.3}
\newcommand{\tierpubnpower}{87}
\newcommand{\tiersibtau}{+0.0221}
\newcommand{\tiersibdz}{0.68}
\newcommand{\tiersibauc}{0.745}
\newcommand{\tiersibtpr}{3.7}
\newcommand{\tiersibnpower}{14}
\newcommand{\tierorctau}{+0.0221}
\newcommand{\tierorcdz}{0.71}
\newcommand{\tierorcauc}{0.767}
\newcommand{\tierorctpr}{3.8}
\newcommand{\tierorcnpower}{13}
\newcommand{\sibshare}{92}
\newcommand{\driftmean}{+0.0001}
\newcommand{\driftsd}{0.0075}
\newcommand{\spillmean}{+0.0243}
\newcommand{\spillsd}{0.0286}
\newcommand{\effsd}{0.0309}
\newcommand{\driftvarshare}{6}
\newcommand{\auczlib}{0.517}
\newcommand{\tprzlib}{1.2}
\newcommand{\aucloss}{0.519}
\newcommand{\tprloss}{1.4}
\newcommand{\aucppl}{0.519}
\newcommand{\tprppl}{1.4}
\newcommand{\aucbpb}{0.519}
\newcommand{\tprbpb}{1.2}
\newcommand{\aucmink}{0.527}
\newcommand{\tprmink}{1.3}
\newcommand{\aucminkpp}{0.559}
\newcommand{\tprminkpp}{1.5}
\newcommand{\aucsurprise}{0.566}
\newcommand{\tprsurprise}{1.6}
\newcommand{\auczstage}{0.562}
\newcommand{\tprzstage}{2.0}
\newcommand{\aucrstage}{0.603}
\newcommand{\tprrstage}{1.3}
\newcommand{\aucminkpprstage}{0.608}
\newcommand{\tprminkpprstage}{1.6}
\newcommand{\aucminkrstage}{0.626}
\newcommand{\tprminkrstage}{1.6}
\newcommand{\aucminkpprsibling}{0.641}
\newcommand{\tprminkpprsibling}{1.2}
\newcommand{\auczsibling}{0.725}
\newcommand{\tprzsibling}{8.1}
\newcommand{\aucrsibling}{0.745}
\newcommand{\tprrsibling}{3.7}
\newcommand{\aucminkrsibling}{0.779}
\newcommand{\tprminkrsibling}{5.3}
\newcommand{\aucminkpproracle}{0.665}
\newcommand{\tprminkpproracle}{1.6}
\newcommand{\aucroracle}{0.767}
\newcommand{\tprroracle}{3.8}
\newcommand{\auczoracle}{0.778}
\newcommand{\tprzoracle}{8.8}
\newcommand{\aucminkroracle}{0.816}
\newcommand{\tprminkroracle}{5.0}
\newcommand{\bestfreeauc}{0.566}
\newcommand{\bestfreename}{Surprise}
\newcommand{\doseexponent}{1.00}
\newcommand{\doseexplo}{0.96}
\newcommand{\doseexphi}{1.04}
\newcommand{\dosertwo}{0.991}
\newcommand{\dosemuone}{+0.0232}
\newcommand{\dosetprone}{6}
\newcommand{\dosedzone}{0.61}
\newcommand{\dosenone}{1982}
\newcommand{\doseaucone}{0.718}
\newcommand{\dosemutwo}{+0.0396}
\newcommand{\dosetprtwo}{14}
\newcommand{\dosedztwo}{0.88}
\newcommand{\dosentwo}{102}
\newcommand{\doseauctwo}{0.844}
\newcommand{\dosemufour}{+0.0802}
\newcommand{\dosetprfour}{41}
\newcommand{\dosedzfour}{1.50}
\newcommand{\dosenfour}{100}
\newcommand{\doseaucfour}{0.934}
\newcommand{\dosemueight}{+0.1593}
\newcommand{\dosetpreight}{79}
\newcommand{\dosedzeight}{2.21}
\newcommand{\doseneight}{99}
\newcommand{\doseauceight}{0.995}
\newcommand{\dosemuonesix}{+0.3281}
\newcommand{\dosetpronesix}{100}
\newcommand{\dosedzonesix}{3.03}
\newcommand{\dosenonesix}{99}
\newcommand{\doseauconesix}{1.000}
\newcommand{\dosemuthreetwo}{+0.7258}
\newcommand{\dosetprthreetwo}{100}
\newcommand{\dosedzthreetwo}{3.40}
\newcommand{\dosenthreetwo}{40}
\newcommand{\doseaucthreetwo}{1.000}
\newcommand{\driftarxivhold}{+0.0421}
\newcommand{\driftarxivtok}{60.8}
\newcommand{\driftarxivnull}{+0.0440}
\newcommand{\driftwikipediahold}{+0.0107}
\newcommand{\driftwikipediatok}{1.6}
\newcommand{\driftwikipedianull}{+0.0113}
\newcommand{\driftstackexchangehold}{+0.0066}
\newcommand{\driftstackexchangetok}{0.8}
\newcommand{\drifteditedwikipediahold}{+0.0044}
\newcommand{\drifteditedwikipediatok}{3.9}
\newcommand{\drifteditedwikipedianull}{+0.0036}
\newcommand{\trajn}{4832}
\newcommand{\trajlag}{495}
\newcommand{\trajpeakzerofour}{0.0843}
\newcommand{\trajendzerofour}{0.0181}
\newcommand{\trajsurvzerofour}{21}
\newcommand{\trajsurvlozerofour}{19}
\newcommand{\trajsurvhizerofour}{24}
\newcommand{\trajprezerofour}{+0.0010}
\newcommand{\trajpeakfoureight}{0.0752}
\newcommand{\trajendfoureight}{0.0220}
\newcommand{\trajsurvfoureight}{29}
\newcommand{\trajsurvlofoureight}{27}
\newcommand{\trajsurvhifoureight}{32}
\newcommand{\trajprefoureight}{-0.0004}
\newcommand{\trajpeakeightonetwo}{0.0694}
\newcommand{\trajendeightonetwo}{0.0231}
\newcommand{\trajsurveightonetwo}{33}
\newcommand{\trajsurvloeightonetwo}{31}
\newcommand{\trajsurvhieightonetwo}{35}
\newcommand{\trajpreeightonetwo}{-0.0011}
\newcommand{\trajpeakonetwoonesix}{0.0581}
\newcommand{\trajendonetwoonesix}{0.0261}
\newcommand{\trajsurvonetwoonesix}{45}
\newcommand{\trajsurvloonetwoonesix}{43}
\newcommand{\trajsurvhionetwoonesix}{47}
\newcommand{\trajpreonetwoonesix}{-0.0005}
\newcommand{\trajpeakonesixtwozero}{0.0425}
\newcommand{\trajendonesixtwozero}{0.0261}
\newcommand{\trajsurvonesixtwozero}{61}
\newcommand{\trajsurvloonesixtwozero}{59}
\newcommand{\trajsurvhionesixtwozero}{63}
\newcommand{\trajpreonesixtwozero}{+0.0007}
\newcommand{\trajpeaktwozerotwothree}{0.0217}
\newcommand{\trajendtwozerotwothree}{0.0178}
\newcommand{\trajsurvtwozerotwothree}{82}
\newcommand{\trajsurvlotwozerotwothree}{79}
\newcommand{\trajsurvhitwozerotwothree}{84}
\newcommand{\trajpretwozerotwothree}{+0.0003}
\newcommand{\decayzeroone}{100}
\newcommand{\decayonetwo}{68}
\newcommand{\decaytwofour}{52}
\newcommand{\decayfoursix}{39}
\newcommand{\decaysixeight}{32}
\newcommand{\decayeightonetwo}{25}
\newcommand{\decayonetwoonesix}{19}
\newcommand{\decayonesixtwozero}{15}
\newcommand{\decaytwozerotwofour}{12}
\newcommand{\seamjump}{-0.00047}
\newcommand{\seammedian}{0.00200}
\newcommand{\seamratio}{0.23}
\newcommand{\seamratiosib}{0.18}
\newcommand{\xsecslopeall}{+0.00004}
\newcommand{\xsecslopepre}{+0.00036}
\newcommand{\xsecci}{[+0.00023,\,+0.00052]}
\newcommand{\xsecn}{4391}
\newcommand{\ptmidtau}{+0.0221}
\newcommand{\ptmiddz}{0.71}
\newcommand{\ptmidauc}{0.767}
\newcommand{\ptmidtpr}{3.8}
\newcommand{\ptmidsd}{0.0309}
\newcommand{\ptmidpos}{85.4}
\newcommand{\ptsfttau}{+0.0155}
\newcommand{\ptsftdz}{0.65}
\newcommand{\ptsftauc}{0.731}
\newcommand{\ptsfttpr}{3.7}
\newcommand{\ptsftsd}{0.0239}
\newcommand{\ptsftpos}{78.8}
\newcommand{\ptsftcorr}{0.59}
\newcommand{\ptdpotau}{+0.0199}
\newcommand{\ptdpodz}{0.58}
\newcommand{\ptdpoauc}{0.708}
\newcommand{\ptdpotpr}{4.8}
\newcommand{\ptdposd}{0.0343}
\newcommand{\ptdpopos}{74.1}
\newcommand{\ptdpocorr}{0.34}
\newcommand{\ptsftretain}{70}
\newcommand{\ptsftsdretain}{77}
\newcommand{\ptdporetain}{90}
\newcommand{\ptdpotprgain}{+24}
\newcommand{\ptmidtauall}{+0.0335}
\newcommand{\ptmiddzall}{0.45}
\newcommand{\ptmidaucall}{0.793}
\newcommand{\ptmidtprall}{7.3}
\newcommand{\ptmidsdall}{0.0741}
\newcommand{\ptmidposall}{89.1}
\newcommand{\ptsfttauall}{+0.0234}
\newcommand{\ptsftdzall}{0.52}
\newcommand{\ptsftaucall}{0.754}
\newcommand{\ptsfttprall}{8.3}
\newcommand{\ptsftsdall}{0.0454}
\newcommand{\ptsftposall}{81.4}
\newcommand{\ptsftcorrall}{0.90}
\newcommand{\ptdpotauall}{+0.0289}
\newcommand{\ptdpodzall}{0.51}
\newcommand{\ptdpoaucall}{0.741}
\newcommand{\ptdpotprall}{9.2}
\newcommand{\ptdposdall}{0.0570}
\newcommand{\ptdpoposall}{77.0}
\newcommand{\ptdpocorrall}{0.82}
\newcommand{\ptsftretainall}{70}
\newcommand{\ptsftsdretainall}{61}
\newcommand{\ptdporetainall}{86}
\newcommand{\ptdpotprgainall}{+26}
\newcommand{\ptcohmmid}{0.718}
\newcommand{\ptcohmsft}{0.760}
\newcommand{\ptcohmdpo}{0.744}
\newcommand{\ptcohlmid}{0.713}
\newcommand{\ptcohlsft}{0.724}
\newcommand{\ptcohldpo}{0.702}
\newcommand{\ptcohwnewmid}{0.874}
\newcommand{\ptcohwnewsft}{0.732}
\newcommand{\ptcohwnewdpo}{0.694}
\newcommand{\ptcohwgapmid}{0.900}
\newcommand{\ptcohwgapsft}{0.758}
\newcommand{\ptcohwgapdpo}{0.729}
\newcommand{\ptcohssemid}{0.962}
\newcommand{\ptcohssesft}{0.806}
\newcommand{\ptcohssedpo}{0.752}
\newcommand{\ptcohemid}{0.950}
\newcommand{\ptcohesft}{0.825}
\newcommand{\ptcohedpo}{0.754}
\newcommand{\ptcohezeromid}{0.957}
\newcommand{\ptcohezerosft}{0.780}
\newcommand{\ptcohezerodpo}{0.732}
\newcommand{\ppvatfive}{17}
\newcommand{\ppvatfifty}{79}
\newcommand{\ppveightprev}{13}
\newcommand{\ppveightfive}{81}
\newcommand{\trajpreabsmax}{0.0011}
\newcommand{\cohlnamedfree}{0}
\newcommand{\cohltprfree}{0.8}
\newcommand{\cohmnamedfree}{0}
\newcommand{\cohmtprfree}{2.4}
\newcommand{\cohssenamedfree}{0}
\newcommand{\cohssetprfree}{2.2}
\newcommand{\cohwgapnamedfree}{0}
\newcommand{\cohwgaptprfree}{2.5}
\newcommand{\cohwnewnamedfree}{0}
\newcommand{\cohwnewtprfree}{1.4}
\newcommand{\tprstratfree}{2.1}
\newcommand{\namedstratfree}{0}
\newcommand{\namedpooledfree}{0}
\newcommand{\cohlnamedpub}{0}
\newcommand{\cohltprpub}{2.4}
\newcommand{\cohmnamedpub}{0}
\newcommand{\cohmtprpub}{4.9}
\newcommand{\cohssenamedpub}{0}
\newcommand{\cohssetprpub}{0.9}
\newcommand{\cohwgapnamedpub}{0}
\newcommand{\cohwgaptprpub}{1.8}
\newcommand{\cohwnewnamedpub}{0}
\newcommand{\cohwnewtprpub}{1.2}
\newcommand{\tprstratpub}{2.8}
\newcommand{\namedstratpub}{0}
\newcommand{\namedpooledpub}{0}
\newcommand{\cohlnamedsib}{0}
\newcommand{\cohltprsib}{5.2}
\newcommand{\cohmnamedsib}{0}
\newcommand{\cohmtprsib}{5.1}
\newcommand{\cohssenamedsib}{289}
\newcommand{\cohssetprsib}{21.4}
\newcommand{\cohwgapnamedsib}{0}
\newcommand{\cohwgaptprsib}{6.7}
\newcommand{\cohwnewnamedsib}{0}
\newcommand{\cohwnewtprsib}{2.8}
\newcommand{\tprstratsib}{8.2}
\newcommand{\namedstratsib}{289}
\newcommand{\namedpooledsib}{0}
\newcommand{\cohlnamed}{0}
\newcommand{\cohlnamedorc}{0}
\newcommand{\cohltprorc}{2.8}
\newcommand{\cohmnamed}{0}
\newcommand{\cohmnamedorc}{0}
\newcommand{\cohmtprorc}{5.5}
\newcommand{\cohssenamed}{812}
\newcommand{\cohssenamedorc}{812}
\newcommand{\cohssetprorc}{52.0}
\newcommand{\cohwgapnamed}{45}
\newcommand{\cohwgapnamedorc}{45}
\newcommand{\cohwgaptprorc}{12.7}
\newcommand{\cohwnewnamed}{0}
\newcommand{\cohwnewnamedorc}{0}
\newcommand{\cohwnewtprorc}{8.5}
\newcommand{\tprstratorc}{16.5}
\newcommand{\namedstratorc}{857}
\newcommand{\namedpooledorc}{0}
\newcommand{\cohlnamedzorc}{0}
\newcommand{\cohltprzorc}{8.0}
\newcommand{\cohmnamedzorc}{153}
\newcommand{\cohmtprzorc}{12.5}
\newcommand{\cohssenamedzorc}{789}
\newcommand{\cohssetprzorc}{65.9}
\newcommand{\cohwgapnamedzorc}{394}
\newcommand{\cohwgaptprzorc}{50.8}
\newcommand{\cohwnewnamedzorc}{442}
\newcommand{\cohwnewtprzorc}{28.5}
\newcommand{\tprstratzorc}{31.4}
\newcommand{\namedstratzorc}{1778}
\newcommand{\namedpooledzorc}{162}
\newcommand{\cohlnamedzsib}{0}
\newcommand{\cohltprzsib}{6.8}
\newcommand{\cohmnamedzsib}{140}
\newcommand{\cohmtprzsib}{11.0}
\newcommand{\cohssenamedzsib}{470}
\newcommand{\cohssetprzsib}{41.9}
\newcommand{\cohwgapnamedzsib}{238}
\newcommand{\cohwgaptprzsib}{27.2}
\newcommand{\cohwnewnamedzsib}{156}
\newcommand{\cohwnewtprzsib}{15.9}
\newcommand{\tprstratzsib}{20.2}
\newcommand{\namedstratzsib}{1004}
\newcommand{\namedpooledzsib}{142}
\newcommand{\cohlnamedzpub}{0}
\newcommand{\cohltprzpub}{2.8}
\newcommand{\cohmnamedzpub}{0}
\newcommand{\cohmtprzpub}{2.3}
\newcommand{\cohssenamedzpub}{0}
\newcommand{\cohssetprzpub}{4.0}
\newcommand{\cohwgapnamedzpub}{0}
\newcommand{\cohwgaptprzpub}{3.2}
\newcommand{\cohwnewnamedzpub}{0}
\newcommand{\cohwnewtprzpub}{1.5}
\newcommand{\tprstratzpub}{2.6}
\newcommand{\namedstratzpub}{0}
\newcommand{\namedpooledzpub}{0}
\newcommand{\tpronenatlo}{2.9}
\newcommand{\tpronenathi}{4.8}
\newcommand{\powerempirical}{81}
\newcommand{\seedsharetwo}{12}
\newcommand{\ptdpodzretain}{81}
\newcommand{\ptdpoaucretain}{78}
\newcommand{\calibn}{4832}
\newcommand{\calibdev}{946}
\newcommand{\calibnall}{6243}
\newcommand{\calibzsize}{946}
\newcommand{\calibkneg}{48}
\newcommand{\calibdraws}{400}
\newcommand{\calibratioorcauc}{0.767}
\newcommand{\calibratioorcauclo}{0.760}
\newcommand{\calibratioorcauchi}{0.775}
\newcommand{\calibratioorctpr}{3.8}
\newcommand{\calibratioorctprlo}{3.0}
\newcommand{\calibratioorctprhi}{4.8}
\newcommand{\calibratioorcstrat}{16.5}
\newcommand{\calibratioorcaucall}{0.782}
\newcommand{\calibratioorcaucloall}{0.775}
\newcommand{\calibratioorcauchiall}{0.788}
\newcommand{\calibratioorctprall}{5.5}
\newcommand{\calibratioorctprloall}{4.2}
\newcommand{\calibratioorctprhiall}{6.6}
\newcommand{\calibratioorcstratall}{21.2}
\newcommand{\calibrmiaorcauc}{0.770}
\newcommand{\calibrmiaorcauclo}{0.761}
\newcommand{\calibrmiaorcauchi}{0.778}
\newcommand{\calibrmiaorctpr}{3.6}
\newcommand{\calibrmiaorctprlo}{2.7}
\newcommand{\calibrmiaorctprhi}{4.8}
\newcommand{\calibrmiaorcstrat}{16.8}
\newcommand{\calibrmiaorcdauc}{-0.000}
\newcommand{\calibrmiaorcdauclo}{-0.000}
\newcommand{\calibrmiaorcdauchi}{+0.000}
\newcommand{\calibrmiaorcdtpr}{-0.3}
\newcommand{\calibrmiaorcdtprlo}{-0.6}
\newcommand{\calibrmiaorcdtprhi}{+0.0}
\newcommand{\calibrmiaorcaucall}{0.782}
\newcommand{\calibrmiaorcaucloall}{0.775}
\newcommand{\calibrmiaorcauchiall}{0.788}
\newcommand{\calibrmiaorctprall}{4.8}
\newcommand{\calibrmiaorctprloall}{4.0}
\newcommand{\calibrmiaorctprhiall}{6.5}
\newcommand{\calibrmiaorcstratall}{21.0}
\newcommand{\calibrmiaorcdaucall}{-0.000}
\newcommand{\calibrmiaorcdaucloall}{-0.000}
\newcommand{\calibrmiaorcdauchiall}{+0.000}
\newcommand{\calibrmiaorcdtprall}{-0.7}
\newcommand{\calibrmiaorcdtprloall}{-1.4}
\newcommand{\calibrmiaorcdtprhiall}{+0.0}
\newcommand{\calibrmiadeforcauc}{0.521}
\newcommand{\calibrmiadeforcauclo}{0.512}
\newcommand{\calibrmiadeforcauchi}{0.529}
\newcommand{\calibrmiadeforctpr}{1.5}
\newcommand{\calibrmiadeforctprlo}{1.1}
\newcommand{\calibrmiadeforctprhi}{1.9}
\newcommand{\calibrmiadeforcstrat}{1.4}
\newcommand{\calibrmiadeforcdauc}{-0.249}
\newcommand{\calibrmiadeforcdauclo}{-0.262}
\newcommand{\calibrmiadeforcdauchi}{-0.236}
\newcommand{\calibrmiadeforcdtpr}{-2.4}
\newcommand{\calibrmiadeforcdtprlo}{-3.5}
\newcommand{\calibrmiadeforcdtprhi}{-1.3}
\newcommand{\calibrmiadeforcaucall}{0.536}
\newcommand{\calibrmiadeforcaucloall}{0.529}
\newcommand{\calibrmiadeforcauchiall}{0.543}
\newcommand{\calibrmiadeforctprall}{1.9}
\newcommand{\calibrmiadeforctprloall}{1.5}
\newcommand{\calibrmiadeforctprhiall}{2.4}
\newcommand{\calibrmiadeforcstratall}{2.4}
\newcommand{\calibrmiadeforcdaucall}{-0.245}
\newcommand{\calibrmiadeforcdaucloall}{-0.255}
\newcommand{\calibrmiadeforcdauchiall}{-0.237}
\newcommand{\calibrmiadeforcdtprall}{-3.6}
\newcommand{\calibrmiadeforcdtprloall}{-4.7}
\newcommand{\calibrmiadeforcdtprhiall}{-2.2}
\newcommand{\calibzorcauc}{0.778}
\newcommand{\calibzorcauclo}{0.771}
\newcommand{\calibzorcauchi}{0.784}
\newcommand{\calibzorctpr}{8.8}
\newcommand{\calibzorctprlo}{7.1}
\newcommand{\calibzorctprhi}{10.7}
\newcommand{\calibzorcstrat}{31.4}
\newcommand{\calibzorcdauc}{+0.010}
\newcommand{\calibzorcdauclo}{+0.005}
\newcommand{\calibzorcdauchi}{+0.015}
\newcommand{\calibzorcdtpr}{+5.0}
\newcommand{\calibzorcdtprlo}{+3.0}
\newcommand{\calibzorcdtprhi}{+7.0}
\newcommand{\calibzorcaucall}{0.807}
\newcommand{\calibzorcaucloall}{0.802}
\newcommand{\calibzorcauchiall}{0.812}
\newcommand{\calibzorctprall}{11.2}
\newcommand{\calibzorctprloall}{9.0}
\newcommand{\calibzorctprhiall}{13.0}
\newcommand{\calibzorcstratall}{42.3}
\newcommand{\calibzorcdaucall}{+0.026}
\newcommand{\calibzorcdaucloall}{+0.021}
\newcommand{\calibzorcdauchiall}{+0.030}
\newcommand{\calibzorcdtprall}{+5.7}
\newcommand{\calibzorcdtprloall}{+3.9}
\newcommand{\calibzorcdtprhiall}{+7.4}
\newcommand{\calibrmiaaorc}{1}
\newcommand{\calibrmiagammaorc}{1.00}
\newcommand{\calibzTorcauc}{0.772}
\newcommand{\calibzTorctpr}{8.1}
\newcommand{\calibzTorcstrat}{34.5}
\newcommand{\calibrmiaseqorcauc}{0.668}
\newcommand{\calibrmiaseqorctpr}{4.6}
\newcommand{\calibrmiaseqorcstrat}{14.8}
\newcommand{\calibratiosibauc}{0.745}
\newcommand{\calibratiosibauclo}{0.737}
\newcommand{\calibratiosibauchi}{0.753}
\newcommand{\calibratiosibtpr}{3.7}
\newcommand{\calibratiosibtprlo}{2.8}
\newcommand{\calibratiosibtprhi}{4.8}
\newcommand{\calibratiosibstrat}{8.2}
\newcommand{\calibratiosibaucall}{0.756}
\newcommand{\calibratiosibaucloall}{0.750}
\newcommand{\calibratiosibauchiall}{0.762}
\newcommand{\calibratiosibtprall}{5.3}
\newcommand{\calibratiosibtprloall}{4.4}
\newcommand{\calibratiosibtprhiall}{6.5}
\newcommand{\calibratiosibstratall}{15.1}
\newcommand{\calibrmiasibauc}{0.748}
\newcommand{\calibrmiasibauclo}{0.739}
\newcommand{\calibrmiasibauchi}{0.756}
\newcommand{\calibrmiasibtpr}{3.7}
\newcommand{\calibrmiasibtprlo}{2.5}
\newcommand{\calibrmiasibtprhi}{4.8}
\newcommand{\calibrmiasibstrat}{10.0}
\newcommand{\calibrmiasibdauc}{+0.000}
\newcommand{\calibrmiasibdauclo}{-0.000}
\newcommand{\calibrmiasibdauchi}{+0.000}
\newcommand{\calibrmiasibdtpr}{-0.1}
\newcommand{\calibrmiasibdtprlo}{-0.5}
\newcommand{\calibrmiasibdtprhi}{+0.0}
\newcommand{\calibrmiasibaucall}{0.756}
\newcommand{\calibrmiasibaucloall}{0.750}
\newcommand{\calibrmiasibauchiall}{0.762}
\newcommand{\calibrmiasibtprall}{4.8}
\newcommand{\calibrmiasibtprloall}{4.2}
\newcommand{\calibrmiasibtprhiall}{6.4}
\newcommand{\calibrmiasibstratall}{14.6}
\newcommand{\calibrmiasibdaucall}{-0.000}
\newcommand{\calibrmiasibdaucloall}{-0.000}
\newcommand{\calibrmiasibdauchiall}{+0.000}
\newcommand{\calibrmiasibdtprall}{-0.5}
\newcommand{\calibrmiasibdtprloall}{-0.7}
\newcommand{\calibrmiasibdtprhiall}{-0.0}
\newcommand{\calibrmiadefsibauc}{0.521}
\newcommand{\calibrmiadefsibauclo}{0.512}
\newcommand{\calibrmiadefsibauchi}{0.529}
\newcommand{\calibrmiadefsibtpr}{1.5}
\newcommand{\calibrmiadefsibtprlo}{1.1}
\newcommand{\calibrmiadefsibtprhi}{1.9}
\newcommand{\calibrmiadefsibstrat}{1.4}
\newcommand{\calibrmiadefsibdauc}{-0.228}
\newcommand{\calibrmiadefsibdauclo}{-0.241}
\newcommand{\calibrmiadefsibdauchi}{-0.215}
\newcommand{\calibrmiadefsibdtpr}{-2.2}
\newcommand{\calibrmiadefsibdtprlo}{-3.6}
\newcommand{\calibrmiadefsibdtprhi}{-1.1}
\newcommand{\calibrmiadefsibaucall}{0.536}
\newcommand{\calibrmiadefsibaucloall}{0.529}
\newcommand{\calibrmiadefsibauchiall}{0.543}
\newcommand{\calibrmiadefsibtprall}{1.9}
\newcommand{\calibrmiadefsibtprloall}{1.5}
\newcommand{\calibrmiadefsibtprhiall}{2.4}
\newcommand{\calibrmiadefsibstratall}{2.4}
\newcommand{\calibrmiadefsibdaucall}{-0.220}
\newcommand{\calibrmiadefsibdaucloall}{-0.230}
\newcommand{\calibrmiadefsibdauchiall}{-0.211}
\newcommand{\calibrmiadefsibdtprall}{-3.3}
\newcommand{\calibrmiadefsibdtprloall}{-4.6}
\newcommand{\calibrmiadefsibdtprhiall}{-2.2}
\newcommand{\calibzsibauc}{0.725}
\newcommand{\calibzsibauclo}{0.718}
\newcommand{\calibzsibauchi}{0.731}
\newcommand{\calibzsibtpr}{8.1}
\newcommand{\calibzsibtprlo}{6.0}
\newcommand{\calibzsibtprhi}{9.6}
\newcommand{\calibzsibstrat}{20.2}
\newcommand{\calibzsibdauc}{-0.021}
\newcommand{\calibzsibdauclo}{-0.025}
\newcommand{\calibzsibdauchi}{-0.016}
\newcommand{\calibzsibdtpr}{+4.3}
\newcommand{\calibzsibdtprlo}{+1.9}
\newcommand{\calibzsibdtprhi}{+6.0}
\newcommand{\calibzsibaucall}{0.747}
\newcommand{\calibzsibaucloall}{0.742}
\newcommand{\calibzsibauchiall}{0.752}
\newcommand{\calibzsibtprall}{9.6}
\newcommand{\calibzsibtprloall}{8.2}
\newcommand{\calibzsibtprhiall}{11.8}
\newcommand{\calibzsibstratall}{28.9}
\newcommand{\calibzsibdaucall}{-0.009}
\newcommand{\calibzsibdaucloall}{-0.012}
\newcommand{\calibzsibdauchiall}{-0.005}
\newcommand{\calibzsibdtprall}{+4.3}
\newcommand{\calibzsibdtprloall}{+3.1}
\newcommand{\calibzsibdtprhiall}{+5.8}
\newcommand{\calibrmiaasib}{1}
\newcommand{\calibrmiagammasib}{1.00}
\newcommand{\calibzTsibauc}{0.716}
\newcommand{\calibzTsibtpr}{7.2}
\newcommand{\calibzTsibstrat}{20.6}
\newcommand{\calibrmiaseqsibauc}{0.653}
\newcommand{\calibrmiaseqsibtpr}{3.8}
\newcommand{\calibrmiaseqsibstrat}{11.0}
\newcommand{\calibratiopubauc}{0.603}
\newcommand{\calibratiopubauclo}{0.593}
\newcommand{\calibratiopubauchi}{0.613}
\newcommand{\calibratiopubtpr}{1.3}
\newcommand{\calibratiopubtprlo}{0.8}
\newcommand{\calibratiopubtprhi}{1.9}
\newcommand{\calibratiopubstrat}{2.8}
\newcommand{\calibratiopubaucall}{0.597}
\newcommand{\calibratiopubaucloall}{0.589}
\newcommand{\calibratiopubauchiall}{0.605}
\newcommand{\calibratiopubtprall}{1.5}
\newcommand{\calibratiopubtprloall}{1.1}
\newcommand{\calibratiopubtprhiall}{2.3}
\newcommand{\calibratiopubstratall}{5.5}
\newcommand{\calibrmiapubauc}{0.602}
\newcommand{\calibrmiapubauclo}{0.591}
\newcommand{\calibrmiapubauchi}{0.611}
\newcommand{\calibrmiapubtpr}{1.2}
\newcommand{\calibrmiapubtprlo}{0.6}
\newcommand{\calibrmiapubtprhi}{1.7}
\newcommand{\calibrmiapubstrat}{2.2}
\newcommand{\calibrmiapubdauc}{-0.000}
\newcommand{\calibrmiapubdauclo}{-0.000}
\newcommand{\calibrmiapubdauchi}{+0.000}
\newcommand{\calibrmiapubdtpr}{-0.1}
\newcommand{\calibrmiapubdtprlo}{-0.3}
\newcommand{\calibrmiapubdtprhi}{+0.0}
\newcommand{\calibrmiapubaucall}{0.597}
\newcommand{\calibrmiapubaucloall}{0.589}
\newcommand{\calibrmiapubauchiall}{0.605}
\newcommand{\calibrmiapubtprall}{1.5}
\newcommand{\calibrmiapubtprloall}{1.0}
\newcommand{\calibrmiapubtprhiall}{2.2}
\newcommand{\calibrmiapubstratall}{5.4}
\newcommand{\calibrmiapubdaucall}{-0.000}
\newcommand{\calibrmiapubdaucloall}{-0.001}
\newcommand{\calibrmiapubdauchiall}{+0.001}
\newcommand{\calibrmiapubdtprall}{-0.1}
\newcommand{\calibrmiapubdtprloall}{-0.2}
\newcommand{\calibrmiapubdtprhiall}{+0.0}
\newcommand{\calibrmiadefpubauc}{0.520}
\newcommand{\calibrmiadefpubauclo}{0.511}
\newcommand{\calibrmiadefpubauchi}{0.529}
\newcommand{\calibrmiadefpubtpr}{1.2}
\newcommand{\calibrmiadefpubtprlo}{1.0}
\newcommand{\calibrmiadefpubtprhi}{1.9}
\newcommand{\calibrmiadefpubstrat}{1.3}
\newcommand{\calibrmiadefpubdauc}{-0.082}
\newcommand{\calibrmiadefpubdauclo}{-0.091}
\newcommand{\calibrmiadefpubdauchi}{-0.070}
\newcommand{\calibrmiadefpubdtpr}{-0.0}
\newcommand{\calibrmiadefpubdtprlo}{-0.5}
\newcommand{\calibrmiadefpubdtprhi}{+0.8}
\newcommand{\calibrmiadefpubaucall}{0.535}
\newcommand{\calibrmiadefpubaucloall}{0.528}
\newcommand{\calibrmiadefpubauchiall}{0.542}
\newcommand{\calibrmiadefpubtprall}{1.8}
\newcommand{\calibrmiadefpubtprloall}{1.4}
\newcommand{\calibrmiadefpubtprhiall}{2.2}
\newcommand{\calibrmiadefpubstratall}{2.3}
\newcommand{\calibrmiadefpubdaucall}{-0.062}
\newcommand{\calibrmiadefpubdaucloall}{-0.069}
\newcommand{\calibrmiadefpubdauchiall}{-0.054}
\newcommand{\calibrmiadefpubdtprall}{+0.2}
\newcommand{\calibrmiadefpubdtprloall}{-0.5}
\newcommand{\calibrmiadefpubdtprhiall}{+0.8}
\newcommand{\calibzpubauc}{0.562}
\newcommand{\calibzpubauclo}{0.557}
\newcommand{\calibzpubauchi}{0.566}
\newcommand{\calibzpubtpr}{2.0}
\newcommand{\calibzpubtprlo}{1.7}
\newcommand{\calibzpubtprhi}{2.9}
\newcommand{\calibzpubstrat}{2.6}
\newcommand{\calibzpubdauc}{-0.042}
\newcommand{\calibzpubdauclo}{-0.049}
\newcommand{\calibzpubdauchi}{-0.033}
\newcommand{\calibzpubdtpr}{+0.7}
\newcommand{\calibzpubdtprlo}{+0.1}
\newcommand{\calibzpubdtprhi}{+1.7}
\newcommand{\calibzpubaucall}{0.565}
\newcommand{\calibzpubaucloall}{0.561}
\newcommand{\calibzpubauchiall}{0.569}
\newcommand{\calibzpubtprall}{4.2}
\newcommand{\calibzpubtprloall}{3.4}
\newcommand{\calibzpubtprhiall}{4.8}
\newcommand{\calibzpubstratall}{4.0}
\newcommand{\calibzpubdaucall}{-0.032}
\newcommand{\calibzpubdaucloall}{-0.038}
\newcommand{\calibzpubdauchiall}{-0.026}
\newcommand{\calibzpubdtprall}{+2.7}
\newcommand{\calibzpubdtprloall}{+1.6}
\newcommand{\calibzpubdtprhiall}{+3.3}
\newcommand{\calibrmiaapub}{1}
\newcommand{\calibrmiagammapub}{1.00}
\newcommand{\calibzTpubauc}{0.565}
\newcommand{\calibzTpubtpr}{2.2}
\newcommand{\calibzTpubstrat}{2.9}
\newcommand{\calibrmiaseqpubauc}{0.541}
\newcommand{\calibrmiaseqpubtpr}{1.3}
\newcommand{\calibrmiaseqpubstrat}{1.9}
\newcommand{\znullmean}{+3.2}
\newcommand{\znullsd}{3.2}
\newcommand{\znullshare}{48}
\newcommand{\zmemmean}{+6.7}
\newcommand{\calibdevgapratio}{-0.013}
\newcommand{\calibdevgaprmia}{-0.013}
\newcommand{\calibtaumedian}{1.34}
\newcommand{\calibneffshare}{74}
\newcommand{\rmiadevaucathree}{0.529}
\newcommand{\rmiadevaucaone}{0.757}
\newcommand{\auczorcnat}{0.778}
\newcommand{\auczorcnatlo}{0.771}
\newcommand{\auczorcnathi}{0.783}
\newcommand{\tpronezorcnat}{8.8}
\newcommand{\tprpointonezorcnat}{2.94}
\newcommand{\dzzorcnat}{1.04}
\newcommand{\poszorcnat}{89.5}
\newcommand{\npowerzorc}{7}
\newcommand{\tpronezorcnatlo}{7.3}
\newcommand{\tpronezorcnathi}{10.9}
\newcommand{\fdrfivezorc}{162}
\newcommand{\fdrtwentyzorc}{967}
\newcommand{\auczsibnat}{0.725}
\newcommand{\auczsibnatlo}{0.718}
\newcommand{\auczsibnathi}{0.731}
\newcommand{\tpronezsibnat}{8.1}
\newcommand{\tprpointonezsibnat}{2.26}
\newcommand{\dzzsibnat}{0.89}
\newcommand{\poszsibnat}{86.4}
\newcommand{\npowerzsib}{8}
\newcommand{\tpronezsibnatlo}{6.0}
\newcommand{\tpronezsibnathi}{9.5}
\newcommand{\fdrfivezsib}{142}
\newcommand{\fdrtwentyzsib}{797}
\newcommand{\auczpubnat}{0.562}
\newcommand{\auczpubnatlo}{0.557}
\newcommand{\auczpubnathi}{0.566}
\newcommand{\tpronezpubnat}{2.0}
\newcommand{\tprpointonezpubnat}{0.35}
\newcommand{\dzzpubnat}{0.36}
\newcommand{\poszpubnat}{65.0}
\newcommand{\npowerzpub}{45}
\newcommand{\tpronezpubnatlo}{1.6}
\newcommand{\tpronezpubnathi}{2.9}
\newcommand{\fdrfivezpub}{0}
\newcommand{\fdrtwentyzpub}{0}
\newcommand{\tprstratorclo}{13.2}
\newcommand{\tprstratorchi}{19.3}
\newcommand{\tprstratzorclo}{26.1}
\newcommand{\tprstratzorchi}{35.4}
\newcommand{\tprstratsiblo}{6.7}
\newcommand{\tprstratsibhi}{10.8}
\newcommand{\tprstratzsiblo}{17.9}
\newcommand{\tprstratzsibhi}{23.4}
\newcommand{\rmiathrfpr}{1.1}
\newcommand{\rmiathrtpr}{4.0}
\newcommand{\calibnholdout}{4832}
\newcommand{\ratiopooledcaught}{186}
\newcommand{\ratiopooledcaughtm}{153}
\newcommand{\zpooledcaught}{427}
\newcommand{\zpooledcaughtm}{399}
\newcommand{\tokshare}{17}
\newcommand{\toksharearxiv}{9}
\newcommand{\tokshareshort}{45}
\newcommand{\cohdauc}{0.943}
\newcommand{\cohdtau}{+0.1508}
\newcommand{\cohdn}{400}
\newcommand{\cohdtpr}{56.2}
\newcommand{\cohdthreetwoauc}{1.000}
\newcommand{\cohdthreetwotau}{+0.7258}
\newcommand{\cohdthreetwon}{40}
\newcommand{\cohdthreetwotpr}{100.0}
\newcommand{\cohdbauc}{0.921}
\newcommand{\cohdbtau}{+0.0757}
\newcommand{\cohdbn}{60}
\newcommand{\cohdbtpr}{21.7}
\newcommand{\coheauc}{0.950}
\newcommand{\cohetau}{+0.0179}
\newcommand{\cohen}{1495}
\newcommand{\cohetpr}{17.7}
\newcommand{\cohezeroauc}{0.957}
\newcommand{\cohezerotau}{+0.0164}
\newcommand{\cohezeron}{299}
\newcommand{\cohezerotpr}{36.5}
\newcommand{\cohezerodauc}{0.997}
\newcommand{\cohezerodtau}{+0.0669}
\newcommand{\cohezerodn}{200}
\newcommand{\cohezerodtpr}{100.0}
\newcommand{\cohfauc}{0.627}
\newcommand{\cohftau}{+0.0094}
\newcommand{\cohfn}{277}
\newcommand{\cohftpr}{1.1}
\newcommand{\cohlauc}{0.713}
\newcommand{\cohltau}{+0.0264}
\newcommand{\cohln}{250}
\newcommand{\cohltpr}{2.8}
\newcommand{\cohmauc}{0.718}
\newcommand{\cohmtau}{+0.0232}
\newcommand{\cohmn}{1982}
\newcommand{\cohmtpr}{5.5}
\newcommand{\cohsseauc}{0.962}
\newcommand{\cohssetau}{+0.0251}
\newcommand{\cohssen}{1000}
\newcommand{\cohssetpr}{52.0}
\newcommand{\cohwgapauc}{0.900}
\newcommand{\cohwgaptau}{+0.0178}
\newcommand{\cohwgapn}{600}
\newcommand{\cohwgaptpr}{12.7}
\newcommand{\cohwnewauc}{0.874}
\newcommand{\cohwnewtau}{+0.0182}
\newcommand{\cohwnewn}{1000}
\newcommand{\cohwnewtpr}{8.5}
\newcommand{\cohxauc}{0.645}
\newcommand{\cohxtau}{+0.0564}
\newcommand{\cohxn}{200}
\newcommand{\cohxtpr}{12.5}

\newcommand{\screenncal}{3865}
\newcommand{\screennpool}{966}
\newcommand{\screenzfiftyminrej}{10}
\newcommand{\screenzfiftytrue}{54}
\newcommand{\screenzfiftytruelo}{21}
\newcommand{\screenzfiftytruehi}{94}
\newcommand{\screenzfiftyfalse}{3}
\newcommand{\screenztwentytrue}{2}
\newcommand{\screenmzfiftytrue}{48}
\newcommand{\screenstratratiofiftytrue}{617}
\newcommand{\screenstratratiofiftyfalse}{17}
\newcommand{\screenstratratiotwentytrue}{48}
\newcommand{\screendomsseratiofiftytrue}{593}
\newcommand{\screenstratzfiftytrue}{1265}
\newcommand{\screenstratzfiftyfalse}{36}
\newcommand{\screenstratztwentytrue}{170}
\newcommand{\screenstratzfivetrue}{5}
\newcommand{\stratsseshareorc}{65}
\newcommand{\npairspooled}{7326}
\newcommand{\xfittprratio}{3.8}
\newcommand{\xfitfprratio}{1.01}
\newcommand{\xfittprz}{8.8}
\newcommand{\cohmbestfree}{0.636}
\newcommand{\pooledarxivrej}{24.6}
\newcommand{\mvnssepooledrej}{12.0}
\newcommand{\cohfnform}{271}
\newcommand{\cohfplacebo}{0.48}
\newcommand{\cohfplacebopdf}{0.47}
\newcommand{\cohfrho}{0.35}
\newcommand{\cohfrholo}{0.10}
\newcommand{\cohfrhohi}{0.53}
\newcommand{\lennfixed}{1616}
\newcommand{\lenaucsixteen}{0.587}
\newcommand{\lenaucfourk}{0.726}
\newcommand{\lendiff}{+0.14}
\newcommand{\lendifflo}{0.12}
\newcommand{\lendiffhi}{0.16}
\newcommand{\lendifffull}{+0.003}
\newcommand{\lendifffulllo}{-0.002}
\newcommand{\lendifffullhi}{+0.008}
\newcommand{\cohlaucfivetwelve}{0.61}
\newcommand{\cohlaucfourk}{0.80}

\newcommand{\ptinstructtau}{+0.0196}
\newcommand{\ptinstructretain}{89}
\newcommand{\ptinstructdz}{0.55}
\newcommand{\ptinstructauc}{0.697}
\newcommand{\ptinstructtpr}{4.7}
\newcommand{\ptinstructdzretain}{77}
\newcommand{\ptinstructaucretain}{74}
\newcommand{\ptinstructcorr}{0.32}
\newcommand{\ptinstructdzall}{0.50}
\newcommand{\ptinstructtprall}{9.1}
\newcommand{\ptrltwotau}{+0.0197}
\newcommand{\ptcohsseinstruct}{0.745}
\newcommand{\ptcohwgapinstruct}{0.718}
\newcommand{\ptcoheinstruct}{0.739}
\newcommand{\ptcohminstruct}{0.737}

\begin{abstract}
Whether a language model was trained on a document matters for copyright,
privacy, and benchmark contamination. Yet most \gls{MI} evaluations
on language models collect members and non-members after the fact, so success
may reflect distribution shift and failure is inconclusive. 
We re-ran OLMo~2~1B's official 50-billion-token mid-training stage,
replacing selected training chunks with documents from \npairsdesign{}
pairs matched on domain and difficulty. For each pair, a coin flip determined which document was inserted and which was held out.
The run released by the \gls{AI2}
from the same seed saw identical data everywhere else and serves as the
counterfactual. One exposure raises a previously unseen document's likelihood
on average but rarely identifies it. Using the counterfactual as reference,
the reference ratio has an \gls{AUC} of \aucnat{} but, with one threshold across
domains, catches \qty{\tpronenat}{\percent} of members at a \qty{1}{\percent} \gls{FPR}. In contrast, \num{\npower} matched pairs give a paired test \qty{80}{\percent} power at the \qty{5}{\percent}
level to detect training on a collection, given same-domain non-members and a
close reference. A reference trained with another seed recovers $\sibshare\%$
of the counterfactual's \gls{AUC} advantage over chance, whereas
reference-free attacks stay within \num{0.07} of chance when pooled over
domains. The full official post-training recipe, re-run on both models,
leaves \qty{\ptinstructretain}{\percent} of the average effect. Even with a closely matched reference,
substantial document-level variability limits individual membership inference. We will release the
checkpoints, per-token log-probabilities, and the evaluation documents
\ifANON.%
\else%
; the code is available at
\url{https://github.com/ali-bana/OLMo-Twin}.
\fi
\end{abstract}

\begin{IEEEkeywords}
membership inference, training-data provenance, memorization, language models,
privacy auditing
\end{IEEEkeywords}

\glsresetall
\section{Introduction}
\label{sec:intro}

Whether a specific document was used to train a language model has
important implications for copyright and data
provenance~\cite{cooper2025files,lee2024talkin,office2025copyright},
training-data transparency~\cite{parliament2024regulation}, and
benchmark contamination~\cite{golchin2024time,oren2024proving}.
\Gls{MI} attacks attempt to answer this question using
information obtained from a trained model, often through statistics
derived from its token log-probabilities. However, their viability is unsettled: a
steady stream of attack methods for language
models~\cite{carlini2021extracting,shi2024detecting,zhang2025min,xie2024recall}
is met by an equally steady line of work arguing that the attacks do not work, or cannot be shown to work,
on pretraining-scale models and that the benchmarks on which they appear to
work are flawed~\cite{duan2024membership,das2025blind,meeus2025sok,zhang2025membership}.

A central difficulty lies in how these attacks are evaluated.
Many existing benchmarks identify members and non-members using
training-data cutoffs or document publication dates.
These populations can differ in topic, writing style, and other
characteristics unrelated to membership. Consequently, an attack may
distinguish the two groups without identifying whether a document was
actually used for training. Indeed, \citet{das2025blind} show that
classifiers with no access to the target model can outperform reported
\gls{MI} attacks on several published benchmarks. Conversely, poor attack performance does not necessarily imply that
training leaves no detectable evidence of membership, but may
indicate that the evaluated statistics are unable to recover that
evidence. Distinguishing these possibilities requires a setting in
which training-data inclusion is known, and its effect on the model
can be measured.

Previous work has addressed parts of this problem through different
experimental designs. Same-distribution evaluations reduce differences
between member and non-member
populations~\cite{duan2024membership,shi2026olmo}, while controlled
insertion studies establish membership ground truth by introducing
known text into training data, including
canaries~\cite{carlini2019secret}, copyright
traps~\cite{meeus2024copyright}, watermarked
data~\cite{wei2024proving}, and randomized insertions of realistic
text~\cite{wei2025hubble}.
Other approaches estimate counterfactual memorization using models
trained on different data subsets or information from training
checkpoints~\cite{zhang2023counterfactual,lesci2024causal}.
These studies %
leave open important questions: how much does including a real document change a language model trained using an established training recipe, and to what extent can \gls{MI} attacks exploit that change?

We investigate these questions using the official 50-billion-token
mid-training stage of OLMo~2~1B~\cite{olmo2025olmo}.
Starting from the released checkpoint at which this stage begins, we
reproduce the seeded training-data order used by the \gls{AI2} and
construct a second run
in which one document from each of \num{\npairsdesign} matched pairs
is randomly selected for insertion into the training stream.
We use the publicly released same-seed run as our
\emph{exact counterfactual} reference: the training-data sequence
is identical except at the substituted positions.
By comparing the changes in document likelihood between randomly
assigned members and their held-out partners, we estimate the
average membership-associated effect under this intervention.
We then examine how well existing attacks recover this signal,
how the results depend on reference-model quality and document
characteristics, and how membership evidence evolves during
training and subsequent post-training.

\subsection{Challenges and Contributions}
\label{sec:contributions}

Our study addresses four challenges in measuring and interpreting
the evidence that training-data membership leaves in language models.

\paragraph*{C1. A same-seed reference at the scale of an official training stage}

Measuring how much including a document changes a model requires a counterfactual: the same model trained on the same data without that document. Testbeds built for this purpose are often trained at small scale or with simplified setups, and may not capture the training dynamics of real-world recipes. We instead build the counterfactual around a released model: \gls{AI2}'s released mid-training run serves as the control (the \emph{twin}), and we train only the treated model. OLMo's data order is fixed by a seed, so we can reconstruct it and give the treated model the same chunks in the same order, replacing only those assigned to injected documents (\Cref{sec:design-twin}). The control is therefore a run trained with the official recipe, and the two runs see identical training data except at the replaced positions.

\paragraph*{C2. Separating membership from document difficulty and domain shifts}

\Gls{MI} must isolate training inclusion from intrinsic document difficulty~\cite{watson2022importance,carlini2022membership} and domain-level likelihood shifts induced by training on related data. 
To separate these effects, we construct \num{\npairsdesign} document pairs—nearly all real text—matched on cohort, length, and pre-treatment difficulty. In each pair, a coin flip assigns one document to injection and the other to holdout. We define a document's \emph{gain} \effectd{} as its mean per-token log-likelihood difference between the treated model and the twin. 
Because the held-out partner was never trained on, its \emph{domain gain} reflects only how the rest of the intervention and the execution differences moved it.
 The difference between an injected document's gain and its partner's yields its \emph{training gain}; averaging across pairs isolates membership effects from shared domain shifts.
We evaluate thirteen cohorts varying source, length, duplication, timing, and representation (\Cref{tab:cohorts-design}). Our headline population comprises \num{\npairsnat} verified-unseen single-exposure pairs from arXiv, Wikipedia, and StackExchange, screened against both training stages for $\ge 40$-word verbatim overlaps (\Cref{sec:design-verify}).
While this population yields a within-pair effect of \qty{\taunat}{nats/token} (\Cref{sec:results-signal}), never-injected arXiv documents also gain \qty{\driftarxivhold}{nats/token} --- surpassing the within-pair effect (\num{\cohmtau}) of injected arXiv documents. This domain spillover (\Cref{sec:results-spillover}) demonstrates that raw model-to-model likelihood differences cannot measure document-specific membership \emph{without a matched comparison baseline}.

\paragraph*{C3. From measurable effects to membership decisions}

Measurable exposure effects do not necessarily translate into reliable \gls{MI}. We evaluate ranking performance and decisions at low \gls{FPR}, including precision and false-discovery screening under specified prevalences, comparing reference-free attacks against attacks using the published checkpoint, an independently seeded run, or the exact counterfactual (\Cref{sec:background}).

On verified-unseen single-exposure documents evaluated against difficulty-matched non-members, pooled reference-free \gls{AUC} stays within \num{0.07} of chance (\Cref{sec:results-attacks}). Using the exact counterfactual, the reference ratio achieves an \gls{AUC} of \num{\aucnat}, but its \gls{TPR} at pooled \qty{1}{\percent} \gls{FPR} is only \qty{\tpronenat}{\percent}; token-level standardization raises this to \qty{\tpronezorcnat}{\percent}. Calibrating thresholds per domain further improves detection to \qty{\tprstratorc}{\percent} and \qty{\tprstratzorc}{\percent}, respectively (\Cref{sec:results-signal}).

\begin{figure}[t]
  \centering
  \includegraphics[width=\columnwidth]{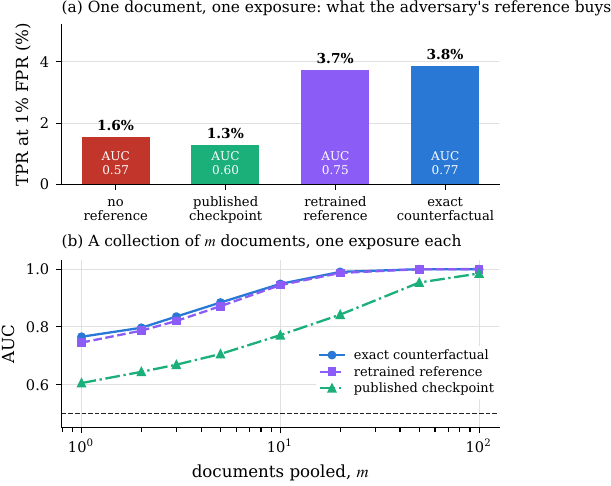}
  \caption{\textbf{One document cannot be decided, but a dozen can.}
  Verified-unseen documents are injected once. (a) \gls{TPR} at \qty{1}{\percent} \gls{FPR}
  with one threshold across domains, for the four reference regimes of
  \Cref{sec:background} (no adversary holds the exact counterfactual); even
  the exact counterfactual catches \qty{\tpronenat}{\percent}
  with the reference ratio and \qty{\tpronezorcnat}{\percent} with the standardized
  ratio (see \S\ref{sec:results-signal}).
  (b) \gls{AUC} when $m$ documents are pooled and compared by their mean score: a
  retrained reference is nearly as good as the counterfactual; at
  $m{=}20$ the counterfactual reaches \num{\collorcmtwozero} and the retrained
  reference \num{\collsibmtwozero}.}
  \label{fig:intro}
\end{figure}

Evidence strengthens considerably when aggregated across documents (\Cref{fig:intro}b). In our paired setup, \num{\npower} matched pairs achieve \qty{80}{\percent} power at the \qty{5}{\percent} level (\Cref{sec:results-signal}), though this assumes access to matched pairs rather than an arbitrary collection audit. Retrained references closely match counterfactual performance: an independently seeded run recovers \qty{\sibshare}{\percent} of the counterfactual's \gls{AUC} advantage over chance (requiring \num{\tiersibnpower} pairs for equal power), while the published checkpoint needs \num{\tierpubnpower} pairs (\Cref{sec:results-tiers}). These results quantify how inference efficacy hinges on reference choice, calibration, and whether the goal is single-document or collection-level detection.

\paragraph*{C4. Duplication, forgetting, and post-training}

Membership evidence depends on exposure timing and frequency. Because early documents are seen at a higher learning rate and are followed by more training, a single final checkpoint obscures exposure-time dynamics. We therefore score every evaluation document under \num{24} step-matched checkpoints of the treated run, the twin and \gls{AI2}'s seed-42069 run. We also carry the treated model and the twin through the official post-training recipe of \gls{SFT}, \gls{DPO}, and two stages of \gls{RLVR}, with identical data, order, and seeds.

For arXiv papers injected up to 32 times, the effect scales nearly linearly with exposure count (fitted exponent \num{\doseexponent}), with per-document detection catching most members at \qty{1}{\percent} \gls{FPR} between four and eight exposures (\Cref{sec:results-dose}). Checkpoint trajectories show that early exposure effects decay substantially during training, despite weak timing dependence at the final checkpoint. Documents injected in the first \num{4000} updates retain $\trajsurvzerofour\%$ of their measured peak at the end of mid-training --- an upper bound, as checkpoints lag injections by \num{\trajlag} updates on average (\Cref{sec:results-forgetting}). After post-training, \qty{\ptinstructretain}{\percent} of the raw within-pair effect and \qty{\ptinstructaucretain}{\percent} of the \gls{AUC} advantage persist at the checkpoint corresponding to the released OLMo~2~1B~Instruct (\Cref{sec:results-posttrain}).

Our results reflect one 1B-parameter model (4T-token pretraining) and one mid-training stage, with \qty{0.135}{\percent} of tokens replaced; \enquote{seen once} denotes a single mid-training injection.
While the reference ratio was pre-specified, all other analyses (standardized ratio, domain thresholds, trajectory, precision, collection-level tests, and post-training comparison) were exploratory post-protocol additions. This work complements critiques of observational \gls{MI}~\cite{meeus2025sok,zhang2025membership} and collection-level dataset inference~\cite{maini2024llm,puerto2025scaling}.

\Cref{sec:background} presents the threat model,
\Cref{sec:design} the experimental design, and
\Cref{sec:methods} the measurement methodology.
\Cref{sec:results} reports the results,
\Cref{sec:discussion} discusses their implications and
limitations, and \Cref{sec:related} reviews related work.

\section{Background and Threat Model}
\label{sec:background}

\paragraph*{Membership inference}
An \gls{MI} attack determines whether a document was used to train a
model~\cite{shokri2017membership}. For language models, common attacks
use token log-probabilities, including loss, zlib-calibrated loss,
Min-K\%, and Min-K\%++~\cite{yeom2018privacy,carlini2021extracting,
shi2024detecting,zhang2025min}. Reference-based attacks instead
compare the target model against a model trained without the document,
helping account for its intrinsic
difficulty~\cite{watson2022importance,carlini2022membership}.
\Cref{sec:related} reviews these approaches, and \Cref{app:attacks}
defines the statistics we evaluate.

\paragraph*{Why the evidence is contested}
Benchmarks that separate members and non-members by publication date or document source may confuse membership with distribution differences. As \citet{das2025blind} show on \foreignlanguage{latin}{post hoc} datasets, simple classifiers without access to the model outperform the best reported attacks. Furthermore, most methods fall somewhere near chance on same-distribution evaluation datasets~\cite{duan2024membership,shi2026olmo}. These concerns motivate evaluations with randomized membership assignment~\cite{meeus2025sok,zhang2025membership}.

\paragraph*{Adversary}
The adversary holds a document $\doc$ and can obtain token-level
log-probabilities from the target model, either through open weights or a suitable \gls{API}. It may additionally hold a reference model, and
the strength of the attack depends on how close that reference is to the
counterfactual \enquote{same model, trained without $\doc$}. We evaluate four
reference regimes (in order of increasing access):
\begin{enumerate}[leftmargin=*, label=\textbf{Regime \arabic*}, ref=Regime \arabic*]\itemsep1pt
  \item \textbf{No reference.} Only the target model.
  \item \textbf{Published checkpoint.} The released checkpoint from which the target's
        final training stage began. It might be available to any adversary today,
        and it has never seen the final stage's data mixture.
  \item\label{itm:retrained} \textbf{Retrained reference.} A model the adversary trains by following the
        published recipe, differing from the target's control only in
        data-order seed. We instantiate it with \gls{AI2}'s own seed-42069 run of
        the same stage, whose data paths are identical to the control's.
  \item \textbf{Exact counterfactual.} The twin \Mctrl. No adversary has this; it
        is the strongest reference in this study.
\end{enumerate}
\ref{itm:retrained} represents a realistic option for an
adversary with access to an open training recipe, although
reproducing the exact data mixture may be difficult.
The exact counterfactual provides the strongest reference
in our study but is not available to a real-world adversary.
\Cref{sec:results-tiers} compares their performance.

\paragraph*{Two questions, and the metrics for each}

\gls{MI} addresses two distinct tasks. The \emph{document-level}
question asks whether a specific document was included in training,
while the \emph{collection-level} question asks whether a set
of documents was included~\cite{maini2024llm,puerto2025scaling}.
The latter allows evidence to be aggregated across documents. Although \gls{AUC} measures overall member vs.\ non-member ranking across all thresholds, document-level claims specifically demand \gls{TPR} at low \gls{FPR}~\cite{carlini2022membership} and positive predictive value at a stated prevalence~\cite{jayaraman2021revisiting}. We report all three metrics and maintain the distinction between these two questions throughout.

\section{A Counterfactual Testbed on an Official Training Run}
\label{sec:design}

\begin{figure}[t]
  \centering
  \includegraphics[clip, trim={0 0 0 2.5cm}, width=\columnwidth]{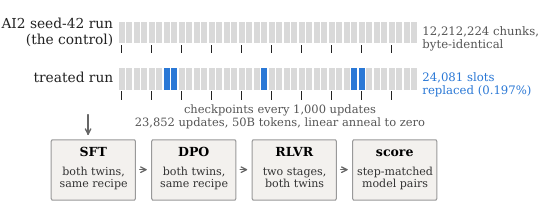}
  \caption{\textbf{The testbed.} \gls{AI2}'s seed-42 run of OLMo~2~1B's stage-2
  mid-training consumes a fixed sequence of \num{12212224} chunks. The treated
  run consumes the same sequence with \num{24081} slots (\qty{0.197}{\percent}) replaced by
  copies of injected documents. Each injected document (a blue slot) has a partner matched
  in length and pre-treatment difficulty that was held out by a coin flip. Both runs save checkpoints every \num{1000} updates (\num{24} per run)
  and both are carried through the official post-training recipe, so
  every document can be scored under step-matched pairs of models at every
  stage.}
  \label{fig:design}
\end{figure}

The testbed has three parts: a treated run that reproduces \gls{AI2}'s official
mid-training stage with documents inserted, a set of injected documents and
their matched partners assigned by coin flip, and the measurements taken on
both models during training and after post-training. \Cref{fig:design} gives the
overview; this section explains each part and how it was verified.

\subsection{The twin}
\label{sec:design-twin}

OLMo~2~1B undergoes a 4-trillion-token pretraining stage followed by
50 billion tokens of mid-training, which anneals the learning rate
linearly to zero~\cite{olmo2025olmo}. The official mid-training stage
runs for \num{23852} updates, each processing \num{512} chunks of
\num{4096} tokens. We start from the released pre-stage
checkpoint, including its optimizer state, and follow the official
mid-training recipe and seed-42 data order.

Our control \Mctrl{} is \gls{AI2}'s released seed-42 run (\gls{AI2} released the training runs using three random seeds, 42, 666, and 42069). We reconstruct
its data order and feed the treated model the same chunks at every
position except the planned substitutions. We verify every inserted chunk against a frozen manifest and
sample unchanged chunks before training and each resumption. An
independent audit after mid-training verifies the data order and the
placement of injected chunks. The training-data sequences therefore match
outside the substitutions.
A separately seeded official run
provides an additional reference (\Cref{sec:results-tiers}), trained on the same data but in a different order.

\subsection{Injection mechanics}
\label{sec:design-injection}

Injection is a substitution, not an addition: each document copy
replaces official training chunks, so both runs process the same
number of tokens. A document of $n$ tokens occupies
$\lceil n/4096\rceil$ consecutive chunk positions, beginning at its
assigned target or the next available position. Any unused space in
the final chunk is filled with tokens from the official chunk it
replaces. All other chunks retain their original contents and order.

Each copy begins at a \num{4096}-token window boundary, matching the
contexts used for scoring (\Cref{sec:methods}). Naturally packed
training documents need not have this alignment, making it a favorable
condition for detection. We distribute placement targets across the
stage, balancing them by training decile and excluding the first and
last \qty{1}{\percent}. Copies of the same document do not share a batch.
In total, \num{12414} copies of \num{\npairsdesign} injected documents
replace \num{24081} chunk positions (\qty{0.197}{\percent} of
positions). The inserted documents contribute
\num{67.3} million tokens, or \qty{0.135}{\percent} of the stream.
The inserted tokens are unevenly distributed across sources; we
examine the resulting domain shifts in \Cref{sec:results-spillover}.

\subsection{Matched pairs assigned by coin flip}
\label{sec:design-pairs}

Candidates come mainly from arXiv papers, Wikipedia pages, and
StackExchange posts. We also include edited versions of previously
trained Wikipedia pages and synthetic anchors, described in \Cref{sec:design-cohorts}.
Within each cohort and stratum, we group candidates by length
quartile, sort them by pre-treatment bits-per-byte under the
pre-stage checkpoint, and pair adjacent documents. An independent
fair coin then selects one document per pair for injection, while
the other is held out. Cohort manipulations, such as exposure count
and target length, apply to both partners.

Matching makes the two arms comparable in source, length, and
pre-treatment difficulty, while random assignment determines which
document is inserted. The consequence is the design's main strength: within a pair, difficulty,
topic, register, and length are controlled by construction, and which partner
becomes the member is chance. A third arm of \num{7053} never-injected documents
from the same sources serves as a background.

\subsection{Cohorts}
\label{sec:design-cohorts}
Inspired by~\cite{bordt2026}, we construct thirteen cohorts to examine document source, length,
exposure count and timing, representation, and novelty (\Cref{tab:cohorts-design}). Our main
\emph{verified-unseen single-exposure} population contains five
cohorts: post-cutoff arXiv papers (M), truncated arXiv prefixes (L),
new Wikipedia pages (W\textsubscript{new}), Wikipedia pages absent
from the training snapshot (W\textsubscript{gap}), and StackExchange
posts (S\textsubscript{se}). This population
contains \num{\npairsnat} matched pairs.

Three cohorts inject arXiv papers
\num{2}--\num{32} times (D, D32, DB), where cohort DB places its four copies inside a
\qty{1}{\percent} window of the run. Cohort F injects the \LaTeX-derived text of a paper
while the partner keeps the PDF-derived text. Cohort X injects synthetic
anchors --- model samples, shuffled text and random tokens --- to span the
surprisal spectrum. The E cohorts are edited Wikipedia pages whose originals
are in both training stages, stratified from \qty{0}{\percent} to over \qty{75}{\percent} novel words;
for these the contrast is re-exposure of known text against known text, not
membership against non-membership, and we report them separately.

\subsection{Verifying non-membership}
\label{sec:design-verify}
Previously seen documents would weaken our member--non-member
comparison. For cohorts intended to be unseen, we therefore search
the candidates against all \num{13} components of the released
pretraining and mid-training data~\cite{olmo2025olmo}, parts of which derive from Dolma~\cite{soldaini2024dolma}.
The exhaustive fingerprint search covers \qty{8.3}{\tebi\byte} and
detects verbatim overlaps of at least \num{40} words. Before training,
we remove \num{156} overlapping candidates and refill their pairs.
Shorter or heavily paraphrased overlaps may remain. The edited
Wikipedia cohorts are intentionally re-exposed, and the synthetic
anchors are not subject to this verification.

\subsection{Post-training}
\label{sec:design-posttrain}

We apply the official post-training recipe to both the
treated model and \Mctrl{}, using identical training data, order,
and seeds.
\gls{SFT} uses the OLMo~2 variant of the T\"ulu~3
mixture~\cite{olmo2025olmo,lambert2025tulu}, followed by \gls{DPO}
on the released preference mixture and two stages of \gls{RLVR}~\cite{lambert2025tulu} as configured in \gls{AI2}'s release of OLMo~2~1B,
the first on math and instruction-following prompts and the second on math
prompts only. \gls{AI2}'s released OLMo~2~1B~Instruct is step~2000 of its second
stage; we report the same step of our runs and the end of the stage. We post-train the control
ourselves rather than compare against \gls{AI2}'s published
instruction-tuned checkpoint. This maintains matched post-training
conditions for measuring how the mid-training intervention persists.

\section{Measurement and Estimation}
\label{sec:methods}

\paragraph*{Scoring}
Every evaluation document is scored from its stored token IDs --- nothing is
re-tokenized --- in non-overlapping \num{4096}-token windows with no cross-window
context, which matches how training saw long documents.
We also record at every position the mean and \gls{sd} of
the log-probability over the vocabulary, which the evaluated attacks require. The
post-trained models are scored on raw document text, not wrapped in a chat
template, to enable comparisons across training stages.
Every checkpoint is scored with the same procedure, token IDs, windows and
positions, and the per-document score is the token-weighted mean over all
scored positions.

\paragraph*{Estimand and estimator}
Let $L_M(\doc)$ denote the mean per-token log-likelihood of
document $\doc$ under model $M$. We define its \emph{gain}
\effectd{} as the difference between the treated and control
models:
\begin{equation}
\effectd =
L_{\mathrm{treat}}(\doc) - L_{\mathrm{ctrl}}(\doc).
\end{equation}
For $N$ matched pairs, we estimate the average
membership-associated effect using the within-pair contrast:
\begin{equation}
\widehat{\tau} =
\frac{1}{N}\sum_{i=1}^{N}
\left[
\effect(\doc_{i,\mathrm{in}})
-
\effect(\doc_{i,\mathrm{out}})
\right].
\end{equation}
The held-out partner captures changes that occur without its
direct inclusion, including shifts shared by related documents.
Random assignment allows the average contrast to estimate the
effect of membership under our joint intervention.
Matching reduces the influence of document difficulty and
shared domain shifts, although these need not cancel exactly
within each pair (\Cref{sec:results-spillover}).
This estimate is not the isolated effect of including one
document while holding all other training data fixed.

\paragraph*{Populations}

Unless stated otherwise, results use the \num{\npairsnat}
verified-unseen single-exposure pairs defined in
\Cref{sec:design-cohorts}. We also report pooled results over
\num{\npairspooled} analyzed pairs from the eligible cohorts,
including duplication and re-exposure experiments.
Cohorts X and F are excluded from pooled estimates and
negative sets: X contains synthetic anchors with unstable
scores across runs, while F differs in surface form between
arms and fails its placebo test (\Cref{sec:results-which}).
Including X and F, \num{\npairs} pairs remain in the full
analyzed dataset after the exclusions for post-training overlap and
document length.

\paragraph*{Inference}
We test the average membership-associated effect using a
sign-flip randomization test based on the within-pair coin
flips. The test is exact under the sharp null of no effect
from assignment. We report Cohen's $d_z$ as a standardized
effect size, bootstrap confidence intervals for means and
\gls{AUC}, and rank \gls{AUC} as a measure of discrimination. 
For individual-document inference, we report \gls{TPR}
at \qty{1}{\percent} and \qty{0.1}{\percent} \glspl{FPR}.
Thresholds are set using held-out partners, either across
the full population (the \emph{pooled} operating point) or
separately for each cohort (the \emph{cohort-stratified}
operating point). The latter requires knowledge of the
document's domain and suitable matched non-members which an attacker might acquire.
We also check the pooled operating point using pair-level
cross-fitting. Where the number of
negatives is too small to resolve an \gls{FPR}, we say which order
statistic set the threshold. To assess which documents can be identified, we construct
candidate pools at stated member prevalences. Each candidate
receives a conformal $p$-value against separate calibration
non-members, followed by the Benjamini--Hochberg procedure
at level $q$. The resulting \gls{FDR} control relies on
assumptions about the calibration and candidate
non-members~\cite{bates2023testing}.
For collection-level inference, sample sizes for
\qty{80}{\percent} power are estimated from the paired
$d_z$ with the normal approximation for a one-sided paired test at
$\alpha=0.05$ and checked by resampling pairs from the observed contrasts.
\Cref{app:stats} provides the statistical details.

\paragraph*{Attacks evaluated}
All statistics are oriented so that higher scores indicate
stronger membership evidence, and all are
computed from the same per-token files. We evaluate mean log-probability (loss),
perplexity, \hbox{bits-per-byte}, the zlib ratio, Min-K\% and Min-K\%++ at
$K{=}20$, and a surprise statistic $\overline{\log p-\mu}$ in which $\mu$ is
the model's own expected log-probability at each position. Reference-based
attacks subtract the same statistic computed under the reference model; for
the mean log-probability this difference is a log-likelihood ratio, and we
call it the \emph{reference ratio}, or \enquote{ratio} in tables.
The pre-specified statistic is the reference ratio; per-cohort numbers use it
rather than the best of the variants. We evaluated $K\in\{5,\dots,50\}$ for
the Min-K\% family and report $K{=}20$; choosing $K$ on half of the pairs and
evaluating on the other half changes the pooled \gls{AUC} by less than \num{0.001}.
A further statistic is computed from the same per-token files
(\Cref{app:attacks}). The \emph{standardized ratio} divides a document's
reference ratio by the standard error of its per-token differences, with the
number of tokens replaced by an effective number $T_{\mathrm{eff}}$ from a windowed
autocorrelation estimate~\cite{sokal1997monte}; it has no tuned parameter.
\Cref{tab:calibrated} compares it with the reference ratio.

\section{Evaluation}
\label{sec:results}

In this section we address eight questions
(\Cref{sec:results-signal,sec:results-attacks,sec:results-tiers,sec:results-forgetting,sec:results-dose,sec:results-spillover,sec:results-posttrain,sec:results-which}). Unless stated otherwise, every number is a
within-pair estimate on verified-unseen documents injected once, measured
against the exact counterfactual. Additionally, all \glspl{CI} are \qty{95}{\percent} \glspl{CI}.

\subsection{Q1: How much signal does one exposure leave, and is it decidable?}
\label{sec:results-signal}

\begin{figure*}[t]
  \centering
  \includegraphics[width=\textwidth]{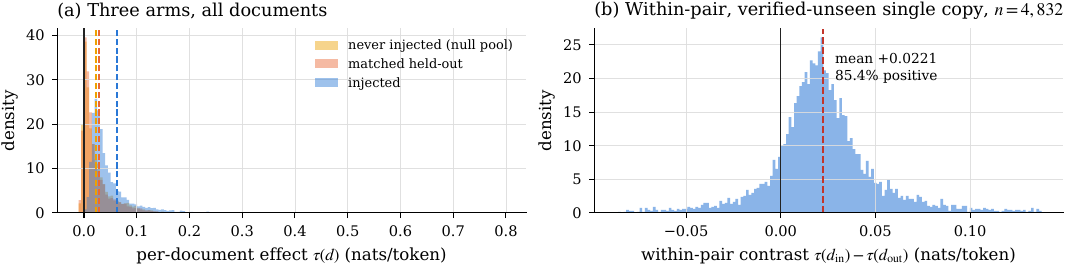}
  \caption{\textbf{One exposure moves a document's likelihood by a small,
  consistent amount.} (a) Per-document effect \effectd{} for the three arms over
  all documents; the never-injected arms are not centered at zero
  (\S\ref{sec:results-spillover}). (b) The within-pair contrast for
  verified-unseen single-exposure pairs removes that offset: the mean is
  $\taunat$ and $\posnat\%$ of pairs are ordered correctly.}
  \label{fig:signal}
\end{figure*}

\begin{figure*}[t]
  \centering
  \includegraphics[width=\textwidth]{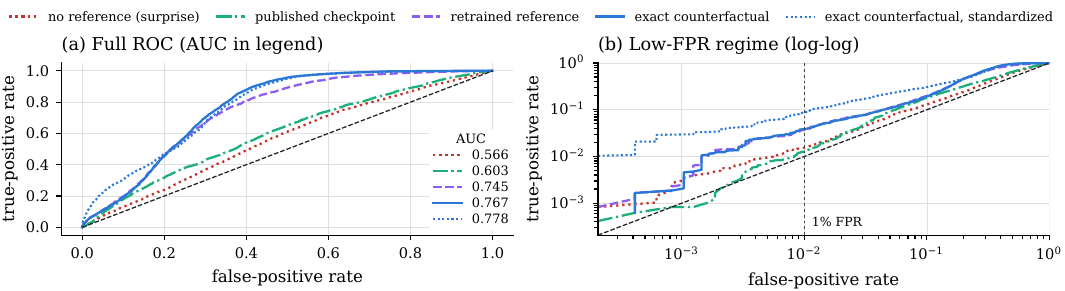}
  \caption{\textbf{No reference regime yields a usable per-document operating
  point for most documents.} \Glsxtrfull{ROC} curves for the four reference regimes, and for the
  standardized ratio against the exact counterfactual (blue dotted), on
  verified-unseen single-exposure pairs, (a) linear and (b) log--log with the
  \qty{1}{\percent} \gls{FPR} line. Standardizing lifts the low-\gls{FPR} end; the other curves converge
  on the diagonal below about \qty{0.1}{\percent} \gls{FPR}.}
  \label{fig:roc}
\end{figure*}

\paragraph*{The effect exists and is unambiguous in aggregate}
Over the \num{\npairsnat} verified-unseen single-exposure pairs, the within-pair
effect is \qty{\taunat}{nats/token} (standard error \num{\senat}; bootstrap %
\gls{CI}
$[\taunatlo, \taunathi]$), with $d_z=\dznat$. The injected document has a higher gain than its held-out partner in
\qty{\posnat}{\percent} of pairs (\Cref{fig:signal}b). The sign-flip
randomization test places the observed mean \num{\permz} \glspl{sd}
outside its own null, so $p\permp$ is the resolution of the test rather than an
estimate. Pooled over the \num{\npairspooled} analyzed pairs of every cohort except F and X,
which include the duplication and re-exposure cohorts, the effect is \num{\tauall} with
$d_z=\dzall$; the standardized effect is smaller there because the multi-copy
cohorts add between-dose variance, not because the signal is weaker.

\begin{figure*}[t]
  \centering
  \includegraphics[width=\textwidth]{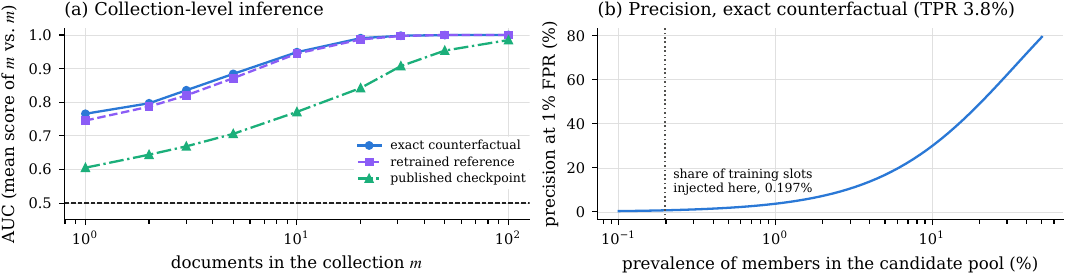}
  \caption{\textbf{Set membership is easy; document membership is not.}
  (a) \gls{AUC} when $m$ documents are pooled and compared by their mean score, for
  three reference regimes. (b) Precision at \qty{1}{\percent} \gls{FPR} as a function of the
  prevalence of members in the candidate pool, at the exact counterfactual's
  \gls{TPR}; at a prevalence of \qty{0.2}{\percent}, about the share of training slots
  injected here, it is below \qty{1}{\percent}.}
  \label{fig:collection}
\end{figure*}

\paragraph*{It does not support a decision about one document}
With the exact counterfactual as reference, \gls{AUC} is \num{\aucnat} (%
\gls{CI}
$[\aucnatlo,\aucnathi]$), but with one threshold for the whole population the
\gls{TPR} at \qty{1}{\percent} false positives is \qty{\tpronenat}{\percent} (bootstrap %
\gls{CI}
\qtyrange[range-units=single,range-phrase=--]{\tpronenatlo}{\tpronenathi}{\percent}) and at \qty{0.1}{\percent} it is \qty{\tprpointonenat}{\percent}
(\Cref{fig:roc}). Calibrating the threshold on separate pairs through
five-fold cross-fitting gives \qty{\xfittprratio}{\percent} at a realized
\gls{FPR} of \qty{\xfitfprratio}{\percent}. Combining the ratio with
Min-K\% improves \gls{AUC} to \num{\aucminkroracle} and the \qty{1}{\percent}
\gls{FPR} detection rate to \qty{\tprminkroracle}{\percent}. The reason is visible in \Cref{fig:signal}b: the contrast
distribution is a broad shift of the whole population, with \gls{sd}
\num{\sdnat} versus a mean of \num{\taunat}, rather than a separated subpopulation of
exposed documents. Precision also depends on how common members are in the population
being screened. At an illustrative prevalence of \qty{0.2}{\percent},
close to the fraction of training slots replaced here, the positive
predictive value at \qty{1}{\percent} \gls{FPR} is only
\qty{\ppvprev}{\percent}: more than \num{99} in \num{100} flagged
documents would be non-members. At \qty{5}{\percent} prevalence
it is \qty{\ppvatfive}{\percent} and at \qty{50}{\percent} it is \qty{\ppvatfifty}{\percent}
(\Cref{fig:collection}b). Naming individual documents
under false-discovery control requires a candidate pool that also contains
non-members. We form such pools at stated member prevalences from held-out
folds of the pairs, compare every candidate with \num{\screenncal} non-members of
the other folds, and apply the Benjamini--Hochberg procedure at $q=0.05$
(\Cref{app:stats}). With the reference ratio, no member is named at any
prevalence from \qty{0.2}{\percent} to \qty{50}{\percent}.

\paragraph*{A per-document variance estimate helps, at the tail}
The reference ratio averages token-level differences, so its precision
varies across documents. Longer documents generally provide a more precise
average than short posts; token-level noise accounts for
\qty{\tokshareshort}{\percent} of the within-pair variance in the short
cohorts (\Cref{sec:results-which}). The \emph{standardized ratio} divides
each document's ratio by its estimated standard error, using an effective
token count to account for autocorrelation (\Cref{sec:methods}). It has
no tuned parameter. Against the exact counterfactual, \gls{AUC} changes
little (\num{\auczorcnat} versus \num{\aucnat}), but \gls{TPR} at
\qty{1}{\percent} \gls{FPR} rises from \qty{\tpronenat}{\percent} to
\qty{\tpronezorcnat}{\percent} (\gls{CI}
\qtyrange[range-units=single,range-phrase=--]{\tpronezorcnatlo}{\tpronezorcnathi}{\percent};
\qty{\xfittprz}{\percent} with cross-fitted thresholds). At
\qty{0.1}{\percent} \gls{FPR}, it rises from
\qty{\tprpointonenat}{\percent} to
\qty{\tprpointonezorcnat}{\percent}. Under false-discovery control, the gain is real but
small: in pools that are half members, the standardized ratio names
\num{\screenzfiftytrue} of the \num{\npairsnat} members on average
(\numrange[range-phrase=--]{\screenzfiftytruelo}{\screenzfiftytruehi} across fold assignments, with
\num{\screenzfiftyfalse} non-members named alongside), about $\screenztwentytrue$
at \qty{20}{\percent} prevalence, and essentially none at \qty{5}{\percent} or below. The
non-members' standardized scores are not centered at zero (mean \num{\znullmean},
SD \num{\znullsd}; the one-sided \qty{1}{\percent} cutoff under a standard-normal
null, $z>2.33$, flags \qty{\znullshare}{\percent} of them), because the treated run moved the held-out partners too
(\Cref{sec:results-spillover}); the statistic is a ranking calibrated against
the matched non-members, like every other statistic here, and not a
$p$-value. Because standardization scales with $\sqrt{T_{\mathrm{eff}}}$, the pooled
threshold favors longer documents. Of the $\zpooledcaught$ members it
detects at \qty{1}{\percent} \gls{FPR}, \num{\zpooledcaughtm} are arXiv
papers, versus \num{\ratiopooledcaughtm} of \num{\ratiopooledcaught}
for the plain ratio. Since pair lengths are matched, this difference
comes from the statistic rather than a length mismatch between arms.

\paragraph*{A threshold per domain changes the answer for short documents}
The pooled threshold is set by the domain whose non-members score highest,
which is arXiv (\Cref{sec:results-spillover}), so members from other domains
are held to a bar set by arXiv papers. An adversary who knows a document's
domain and holds domain-matched non-members can set one threshold per cohort.
At \qty{1}{\percent} \gls{FPR} per cohort the \gls{TPR} weighted by cohort size is
\qty{\tprstratorc}{\percent} (bootstrap %
\gls{CI} \qtyrange[range-units=single,range-phrase=--]{\tprstratorclo}{\tprstratorchi}{\percent};
the per-domain rates are far less stable than the pooled ones because each
cohort's threshold rests on a few negatives), but it is $\cohssetpr\%$ for
StackExchange posts,
\qty{\cohwgaptpr}{\percent} and \qty{\cohwnewtpr}{\percent} for the two Wikipedia cohorts, and
\qty{\cohmtpr}{\percent} and \qty{\cohltpr}{\percent} for arXiv papers and their prefixes
(\Cref{tab:cohorts}). With separate screening within each cohort in pools that are half
members, \num{\screenstratratiofiftytrue} members are named at $q=0.05$ (with
\num{\screenstratratiofiftyfalse} non-members), \num{\screendomsseratiofiftytrue} of
them StackExchange posts, and \num{\screenstratratiotwentytrue} at \qty{20}{\percent}
prevalence. With the standardized ratio the stratified rate is
\qty{\tprstratzorc}{\percent} (%
\gls{CI} \qtyrange[range-units=single,range-phrase=--]{\tprstratzorclo}{\tprstratzorchi}{\percent};
\qty{\cohmtprzorc}{\percent} for arXiv papers), and the within-domain screen names
\num{\screenstratzfiftytrue} members at \qty{50}{\percent} prevalence (with
\num{\screenstratzfiftyfalse} non-members), \num{\screenstratztwentytrue} at \qty{20}{\percent}
and \num{\screenstratzfivetrue} at \qty{5}{\percent}; among arXiv papers it names
\num{\screenmzfiftytrue} of \num{\cohmn} at \qty{50}{\percent} prevalence and none at \qty{5}{\percent} or
below. With the retrained reference, the stratified rate is \qty{\tprstratsib}{\percent},
and with the published checkpoint \qty{\tprstratpub}{\percent}.
Per-document membership inference is therefore not uniformly hard: it
fails for most long natural-text documents and succeeds for short, formulaic
ones, given a reference close to the counterfactual and non-members from the
same domain.

\paragraph*{It does support a decision about a collection}
The same evidence identifies a set easily. For the within-pair contrast
tested against zero with a one-sided paired $t$-test at $\alpha=0.05$, the
observed $d_z$ implies that \num{\npower} matched pairs give \qty{80}{\percent} power, and
resampling \num{\npower} pairs from the observed contrasts rejects in
\qty{\powerempirical}{\percent} of draws; with the standardized ratio the within-pair
$d_z$ is \num{\dzzorcnat} and resampling shows that \num{\npowerzorc} pairs suffice. Pooling $m$
independently drawn members against $m$ non-members and comparing their mean
scores, \gls{AUC} rises from \num{\aucnat} at $m{=}1$ to \num{\collorcmonezero} at
$m{=}10$ and \num{\collorcmtwozero} at $m{=}20$ (\Cref{fig:collection}a); with the
published checkpoint it reaches \num{\collpubmonezero} at $m{=}10$ and needs
\num{\tierpubnpower} pairs for \qty{80}{\percent} power. 
These results do not mean that an arbitrary collection of
\num{\npower} documents can establish membership. The paired test
assumes matched non-members, and a weaker reference requires more
evidence. As we discuss in \Cref{sec:discussion}, domain shifts
(\Cref{sec:results-spillover}) do not average away with collection
size. The difference between evidence for one document and evidence
for a well-matched collection is the central quantitative result
of this paper.

\subsection{Q2: What do existing attacks recover when difficulty is matched?}
\label{sec:results-attacks}

\begin{figure}[bth]
  \centering
  \includegraphics[width=\columnwidth]{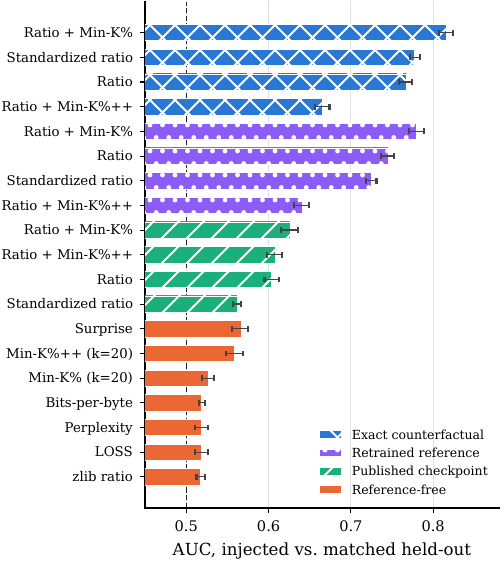}
  \caption{\textbf{Difficulty matching collapses the reference-free statistics we evaluate.}
  \gls{AUC} of every attack statistic on verified-unseen single-exposure pairs with
  bootstrap \qty{95}{\percent} intervals; the dashed line is chance. Pooled over domains, reference-free
  statistics lie within \num{0.07} of chance; what a reference buys depends on how
  close it is to the counterfactual. The standardized ratio changes \gls{AUC}
  little; its gain is at low \gls{FPR} (\Cref{tab:calibrated}).}
  \label{fig:attacks}
\end{figure}

\paragraph*{Reference-free statistics measure difficulty}
\Cref{fig:attacks} and \Cref{tab:attacks} give the central measurement.
Against matched partners, mean log-probability reaches \gls{AUC} $\aucloss$,
Min-K\% \num{\aucmink}, the zlib ratio \num{\auczlib} and bits-per-byte \num{\aucbpb};
the best reference-free statistic, \bestfreename, reaches \num{\bestfreeauc}. The
bootstrap intervals of several exclude \num{0.5} --- at $n{\approx}$\num{4800} per arm
a margin of \num{0.02} is detectable --- so the honest statement is not that these
attacks fail entirely but that they retain a statistically visible and
operationally worthless margin: none exceeds \qty{1.6}{\percent} \gls{TPR} at \qty{1}{\percent} \gls{FPR}. We read
this as evidence about what the statistics measure. They rank documents mainly by
how surprising they are to the model; members and non-members in a \foreignlanguage{latin}{post hoc} split
differ in surprisingness, matched pairs do not, and when the difficulty axis
is removed, little remains. This is consistent with \citet{das2025blind},
who outperform the best reported attacks on \foreignlanguage{latin}{post hoc}
benchmark sets with no model access at all, which shows that members and
non-members in such sets differ before any model is consulted.

\paragraph*{Position-wise calibration appears more useful than token selection}
Min-K\%++ reaches \num{\aucminkpp} against \hbox{Min-K\%'s} \num{\aucmink}. The surprise
statistic --- every token calibrated by the model's own expected log-probability
at that position, with no selection of least-likely tokens --- reaches
\num{\aucsurprise}, slightly more. The gain therefore comes from per-position
calibration rather than from reading only the hardest tokens, which is a
testable claim about \hbox{Min-K\%++} that, to our knowledge, has not previously been
isolated. However, the comparison is not conclusive; the intervals
overlap and the surprise statistic also drops \hbox{Min-K\%++'s} division by the
\gls{sd}. OLMo-Detect~\cite{shi2026olmo} reports the opposite
ordering on sets matched on quality, date and vocabulary but not on difficulty
(\hbox{Min-K\%++} \num{0.56} against \hbox{Min-K\%'s} \num{0.66}, averaged over model sizes
and training stages); in our data the gain comes mainly from the arXiv papers
of cohort M (\num{0.613} against \num{0.537}; the surprise statistic reaches
$\cohmbestfree$ there, the highest reference-free \gls{AUC} among the
single-exposure cohorts), while on StackExchange posts the two are close
(\num{0.525} against \num{0.513}). With a reference, the pattern reverses: adding \hbox{Min-K\%} to
the counterfactual ratio raises \gls{AUC} from \num{\aucroracle} to \num{\aucminkroracle},
while adding \hbox{Min-K\%++} lowers it to \num{\aucminkpproracle}.

\paragraph*{Offline LiRA reduces to the ratio; the standardized ratio helps at the tail}
The likelihood-ratio attack (LiRA)~\cite{carlini2022membership}
in its online form needs reference models trained with the document, which no
run here provides. Its offline form needs only models trained without the
document, but estimates each document's score distribution under them from
many such models; with one reference model and a variance shared across
documents, offline LiRA on the mean log-likelihood ranks documents exactly as
the reference ratio does, so we do not evaluate it separately. On the
\num{\npairsnat} verified-unseen single-exposure pairs (\Cref{tab:calibrated}),
the standardized ratio raises the \gls{TPR} at \qty{1}{\percent} \gls{FPR} from \qty{\calibratioorctpr}{\percent}
to \qty{\calibzorctpr}{\percent} (paired difference \num{\calibzorcdtpr} points, %
\gls{CI}
$[\calibzorcdtprlo,\calibzorcdtprhi]$) at an \gls{AUC} change of \num{\calibzorcdauc}
$[\calibzorcdauclo,\calibzorcdauchi]$; with the retrained reference from
\qty{\calibratiosibtpr}{\percent} to \qty{\calibzsibtpr}{\percent} at an \gls{AUC} change of
\num{\calibzsibdauc}; and with the published checkpoint it gains little. Its
advantage is at the tail, where the documents whose ratio is large relative
to their own token noise sit; the bulk of the ranking, which \gls{AUC} measures, is
governed by the document-level dispersion of \Cref{sec:results-tiers}, which
no per-document variance estimate removes.

\subsection{Q3: How good a reference does the adversary need?}
\label{sec:results-tiers}
\begin{figure}[t]
  \centering
  \includegraphics[clip, trim={8.6cm 0 0 0.3cm}, width=\columnwidth]{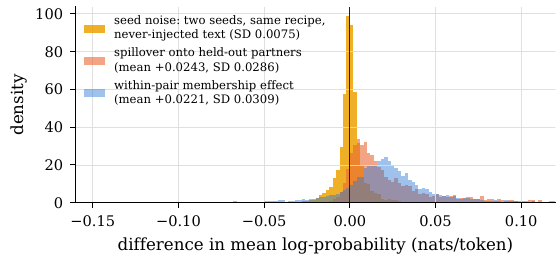}
  \caption{\textbf{A retrained reference is almost as good as the exact
  counterfactual, because seed noise is small.}
  Three distributions over
  verified-unseen documents: the difference between two seeds of the same
  recipe on never-injected text (seed noise), the difference the injection
  made to the held-out partners (spillover, \S\ref{sec:results-spillover}),
  and the within-pair membership effect; about \qty{1}{\percent} of the spillover and
  membership values lie beyond the axis and are not drawn.}
  \label{fig:tiers}
\end{figure}

\paragraph*{The published checkpoint is a weak reference while a retrained reference recovers most of the counterfactual's advantage}
Differencing against the pre-stage-2 checkpoint lifts \gls{AUC} from $\aucloss$ to
$\tierpubauc$ but $d_z$ only to \num{\tierpubdz}, so a collection-level test with
this reference needs \num{\tierpubnpower} pairs for \qty{80}{\percent} power.
Against AI2's seed-42069 run --- the same recipe and the same data mixture as
the control, a different random order --- the within-pair effect is
\num{\tiersibtau} with $d_z=\tiersibdz$ and \gls{AUC} \num{\tiersibauc}, versus
\num{\tierorctau}, \num{\tierorcdz}, and \num{\tierorcauc} for the exact counterfactual
(\Cref{tab:attacks}). The effect estimate is unchanged; the retrained
reference recovers \qty{\sibshare}{\percent} of the counterfactual's advantage over
chance; and a collection-level test needs \num{\tiersibnpower} rather than
\num{\tierorcnpower} pairs. \Cref{fig:tiers} explains why. On never-injected
text, two seeds of the same recipe differ by \qty{\driftmean}{nats/token} on
average with \gls{sd} \num{\driftsd}; with both members of a pair
carrying that noise, it is about \qty{\seedsharetwo}{\percent} of the variance of the
within-pair effect (\gls{sd} $\effsd$). The dispersion an attack must clear is far
larger: the treated run moved its never-injected held-out partners by
\qty{\spillmean}{nats/token} on average with \gls{sd} \num{\spillsd}. That
spread is how unevenly the injection as a whole, together with any
hardware nondeterminism we cannot separate from it, moved documents the
treated model never saw; a single-document counterfactual would not contain
it, and it is what the per-document results of this paper are measured
against.

A caveat bounds this result. The seed-42069 run shares the control's exact
data mixture rather than one the adversary reconstructed from a paper, so the
retrained-reference numbers are themselves an upper bound on that tier. What
the regime establishes is that the seed is not the obstacle, not that mixture
recovery is free.

\subsection{Q4: Is the signal forgotten?}
\label{sec:results-forgetting}

\begin{figure*}[t]
  \centering
  \includegraphics[width=\textwidth]{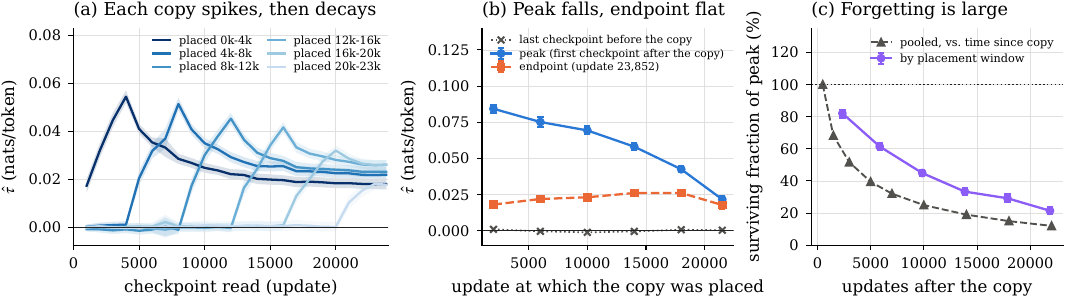}
  \caption{\textbf{Forgetting is large; the annealed learning rate hides it.}
  All panels use verified-unseen single-exposure pairs scored under \num{24}
  step-matched checkpoints. (a) The within-pair effect as a function of the
  checkpoint it is read at, grouped by when the copy was placed: each curve is
  flat at zero until its own copy lands, then spikes and decays. (b) Aligned to
  each document's own placement: the peak falls with placement time because
  the learning rate anneals, the endpoint barely moves, and the last
  checkpoint before placement is zero. (c) The fraction of the peak surviving
  to the end of the stage against how much training followed, by placement
  window and pooled against time since the copy.}
  \label{fig:trajectory}
\end{figure*}

The answer is that most of it is forgotten, and that the end-of-training
snapshot cannot show it. We first state what the snapshot says.

\paragraph*{The cross-section says nothing is forgotten}
Regressing the end-of-training effect on the update at which the copy was
placed gives a slope of \qty{\xsecslopepre}{nats/token} per \num{1000} updates over
the first \qty{90}{\percent} of the run (bootstrap CI $\xsecci$, $n{=}\xsecn$): flat to
slightly increasing. A document injected at update \num{2000} is as detectable at
the end as one injected at update \num{20000}, and the natural reading is that a
single exposure is not forgotten over 45 billion subsequent tokens. That
reading is wrong, and the cross-section cannot show why, because in it two
things are perfectly collinear: a document injected late was both seen
\emph{recently} and seen at a \emph{low learning rate}.

\paragraph*{Along a trajectory the two separate}
We scored the whole evaluation set under every checkpoint (one per \num{1000} updates) of the
treated run, the control and the seed-42069 run (\num{69} checkpoints), and aligned
each document to the first checkpoint at or after its own placement.
\Cref{fig:trajectory}a shows the result: every curve is flat at zero until its
copy lands --- the last checkpoint before placement is within
$\pm\trajpreabsmax$ nats/token of zero in every window, a placebo the design
did not plan --- then spikes and decays. Both models consume identical data
after the copy, so the decline is not the control catching up on the domain;
it is the treated model's trace of the copy being overwritten. Placement time
also fixes a document's position in the data mixture, which we cannot separate
from the learning rate at exposure.
\Cref{fig:trajectory}b separates the height of the peak from how much of it
survives. The peak falls with placement time, from \num{\trajpeakzerofour} for
copies placed before update \num{4000} to \num{\trajpeaktwozerotwothree} for those
placed after \num{20000}, because the learning rate anneals linearly to zero. The
fraction of the peak surviving to the end rises in the opposite direction,
from $\trajsurvzerofour\%$ (%
\gls{CI} \numrange[range-phrase=--]{\trajsurvlozerofour}{\trajsurvhizerofour})
to \qty{\trajsurvtwozerotwothree}{\percent}, because less training follows
(\Cref{fig:trajectory}c). Pooled against time since the copy, the effect
retains \qty{\decayonetwo}{\percent} of its peak after \numrange[range-phrase=--]{1000}{2000} updates,
$\decayfoursix\%$ after \numrange[range-phrase=--]{4000}{6000} and \qty{\decayonetwoonesix}{\percent} after
\numrange[range-phrase=--]{12000}{16000}. The endpoint is nearly flat in placement time only because
these two gradients nearly cancel.

Two qualifications apply. Checkpoints are \num{1000} updates apart, so the first
reading after a copy is on average \num{\trajlag} updates late and the true peak,
immediately after the gradient step, is higher; every surviving fraction above
is an upper bound on retention. And the cancellation is a property of this
learning-rate schedule: under a schedule that anneals differently, the
cross-section would not be flat, which is a reason not to extrapolate the
\enquote{no forgetting} reading beyond runs like this one.

\subsection{Q5: How does the signal scale with duplication?}
\label{sec:results-dose}

\begin{figure}[t]
  \centering
  \includegraphics[width=\columnwidth]{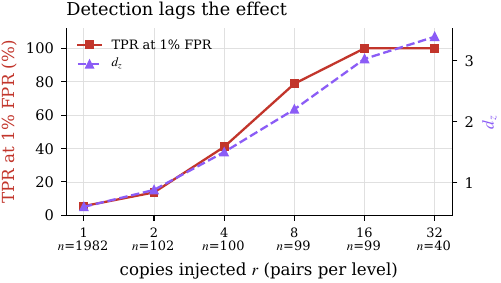}
  \caption{\textbf{Detection lags the effect.} True-positive rate at \qty{1}{\percent}
  \gls{FPR} and $d_z$ against the number of copies $r$ on the matched arXiv
  ladder (cohort M at one copy, D at 2--16, D32 at 32); the effect itself grows
  linearly in $r$. At $r{\ge}8$ the \qty{1}{\percent} \gls{FPR} threshold is the single largest
  held-out score, so those points sit at the resolution floor of
  $n{\approx}100$ negatives ($40$ at $r{=}32$).}
  \label{fig:dose}
\end{figure}
On the matched arXiv cohorts, the average membership effect
grows nearly linearly with the number of exposures $r$.
A power-law fit gives $r^{\doseexponent}$ (\gls{CI}
\numrange[range-phrase=--]{\doseexplo}{\doseexphi},
$R^2=\dosertwo$). The measured effects are
\num{\dosemuone}, \num{\dosemutwo}, \num{\dosemufour},
\num{\dosemueight}, \num{\dosemuonesix}, and
\num{\dosemuthreetwo} nats/token for
$r=1,2,4,8,16,32$, respectively. We observe no
clear saturation over this range. This is consistent with
the increased extraction and regeneration rates under
duplication reported by \citet{carlini2023quantifying}
and \citet{kandpal2022deduplicating}, although those
studies measure different outcomes.
Individual detection improves more slowly because
the variation across documents also increases with
exposure count. At $r=1,2,4,8$, Cohen's $d_z$ is
\num{\dosedzone}, \num{\dosedztwo}, \num{\dosedzfour},
and \num{\dosedzeight}, while \gls{TPR} at
\qty{1}{\percent} \gls{FPR} rises from
\qty{\dosetprone}{\percent} to
\qty{\dosetprtwo}{\percent},
\qty{\dosetprfour}{\percent}, and
\qty{\dosetpreight}{\percent} (\Cref{fig:dose}).
Detection of most members therefore becomes possible
between four and eight exposures in this setting.
However, precision still depends on member prevalence:
at $r=8$, it is only \qty{\ppveightprev}{\percent}
at \qty{0.2}{\percent} prevalence, compared with
\qty{\ppveightfive}{\percent} at
\qty{5}{\percent} prevalence.
These low-\gls{FPR} estimates have limited resolution.
The duplication cohorts contain approximately
\num{100} pairs per exposure level, so the
\qty{1}{\percent} \gls{FPR} threshold is determined
by only one or two held-out scores. At $r\ge8$,
it is set by the single largest held-out score.
The high detection rates should therefore be
interpreted as evidence of strong detectability,
rather than precise estimates of performance at
\qty{1}{\percent} \gls{FPR}.

\subsection{Q6: Do documents the model never saw also move?}
\label{sec:results-spillover}

\begin{figure}[t]
  \centering
  \includegraphics[width=\columnwidth]{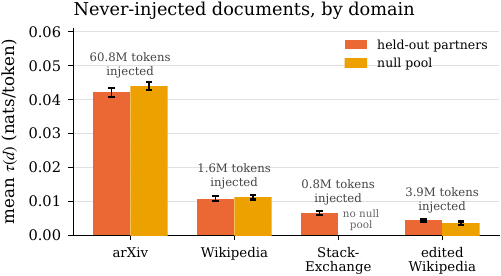}
  \caption{\textbf{Documents in neither model still gain.} Mean \effectd{} for
  never-injected documents by domain --- the held-out partners and, except for
  StackExchange, an independent null pool --- with the number of tokens injected in that domain.
  The domain that received \qty{90}{\percent} of the injected tokens moves most; the
  ordering among the small domains does not follow token share.}
  \label{fig:drift}
\end{figure}

The never-injected documents in \Cref{fig:signal}a are
not centered at zero: their likelihoods also change
between the treated and control models. These shifts are
larger than typical seed-to-seed differences
(\Cref{fig:tiers}) and vary substantially across domains. Never-injected
arXiv papers gain \qty{\driftarxivhold}{nats/token} (held-out partners) and
\num{\driftarxivnull} (null pool); never-injected Wikipedia pages gain
\num{\driftwikipediahold}, StackExchange posts \num{\driftstackexchangehold} and the
edited-Wikipedia pool \num{\drifteditedwikipediahold} (\Cref{fig:drift}). ArXiv
received $\driftarxivtok$M of the $67.3$M injected tokens, so the largest shift
falls on the domain that received the most, but the ordering among the smaller
domains does not follow token share --- edited Wikipedia received more than
plain Wikipedia and moved less --- so we do not claim a dose--response for
spillover, only its existence and its size. 
Its size matters more than it looks. An adversary comparing a target against
any reference without domain-matched non-members measures membership plus this
term, and on arXiv the term (\num{\driftarxivhold}) is \emph{larger} than the
membership effect (\num{\cohmtau}). A population-level claim built on unmatched
non-members can therefore be dominated by the fact that the model saw
\emph{similar} documents, not these ones. Within a pair, both documents share a
domain, so the term cancels in the within-pair contrast on average; testing \effectd{}
against zero instead of against the partner would have inflated our headline
effect by about a factor of two. This spillover is consistent with the finding
of \citet{grosse2023studying} that related training sequences, not only
copies, raise a text's likelihood, and with the contribution of fuzzy
duplicates to a sequence's membership signal~\cite{shilov2026mosaic}, although
that work concerns near-copies rather than distinct documents from one domain.
The spillover is why matched non-members are not a refinement of the design
but its core.

\subsection{Q7: Does post-training erase it?}
\label{sec:results-posttrain}

One might expect post-training to erase evidence of earlier training.
To test this, we apply the same official post-training recipe
(\gls{SFT}, \gls{DPO}, and two stages of \gls{RLVR}), with identical data, order, settings, and seeds,
to both the treated and control models. After \gls{SFT} the within-pair
effect is \num{\ptsfttau} (\qty{\ptsftretain}{\percent} of its mid-training value) and after
\gls{DPO} it is \num{\ptdpotau} (\qty{\ptdporetain}{\percent}); the standardized effect falls from
\num{\ptmiddz} to \num{\ptsftdz} and \num{\ptdpodz}, and \gls{AUC} from \num{\ptmidauc} to
\num{\ptsftauc} and \num{\ptdpoauc}, while the \gls{TPR} at \qty{1}{\percent} \gls{FPR} moves
from \qty{\ptmidtpr}{\percent} to \qty{\ptsfttpr}{\percent} and \qty{\ptdpotpr}{\percent}
(%
\Cref{tab:stages}).
After the two \gls{RLVR} stages, at the checkpoint corresponding to the released
OLMo~2~1B~Instruct (step~2000 of the second stage, to which \enquote{after \gls{RLVR}}
refers below), the within-pair effect is \num{\ptinstructtau} (\qty{\ptinstructretain}{\percent}), with $d_z$
\num{\ptinstructdz}, \gls{AUC} \num{\ptinstructauc} and \gls{TPR} \qty{\ptinstructtpr}{\percent}; at the end of the stage it
is \num{\ptrltwotau}. \gls{RLVR} thus leaves the raw effect nearly where \gls{DPO} left it and
lowers \gls{AUC} by about \num{0.01}.
Post-training therefore does not remove the evidence --- nine tenths of the raw
effect survives the full recipe, \qty{\ptinstructdzretain}{\percent} of the standardized effect
and \qty{\ptinstructaucretain}{\percent} of the \gls{AUC} advantage over chance --- but on
single-exposure documents it does not sharpen it either: the per-pair effect correlates only \num{\ptsftcorr} between
mid-training and \gls{SFT}, \num{\ptdpocorr} between mid-training and \gls{DPO} and \num{\ptinstructcorr} after \gls{RLVR}, so which
documents carry the signal is partly reshuffled. Pooled over all cohorts except F and
X, the picture looks different --- $d_z$ rises from \num{\ptmiddzall} to \num{\ptdpodzall} and
\gls{TPR} from \qty{\ptmidtprall}{\percent} to \qty{\ptdpotprall}{\percent} (\num{\ptinstructdzall} and \qty{\ptinstructtprall}{\percent} after \gls{RLVR}) --- but that rise is carried by the
multi-copy cohorts, whose spread compresses more than their large effects; it is
not a property of the single-exposure case. \gls{DPO} partly reverses \gls{SFT}'s compression. Detectability also falls unevenly
across document types: from mid-training to \gls{DPO}, \gls{AUC} falls from
\num{\ptcohssemid} to \num{\ptcohssedpo} for StackExchange, \num{\ptcohwgapmid} to
\num{\ptcohwgapdpo} for W\textsubscript{gap} and \num{\ptcohemid} to \num{\ptcohedpo} for the
re-exposure cohort E, while arXiv (M) moves from \num{\ptcohmmid} to
\num{\ptcohmdpo}. After \gls{RLVR} these four are \num{\ptcohsseinstruct}, \num{\ptcohwgapinstruct},
\num{\ptcoheinstruct} and \num{\ptcohminstruct}.

\subsection{Q8: Which documents are attackable, and does length help?}
\label{sec:results-which}

\begin{figure}[t]
  \centering
  \includegraphics[width=\columnwidth]{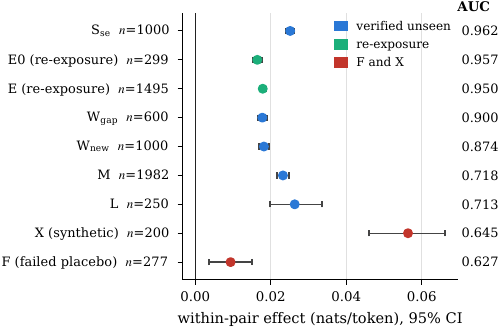}
  \caption{\textbf{Detectability is governed by the variance of the score,
  not only by the size of the effect.} Single-exposure cohorts: within-pair
  effect with \qty{95}{\percent} intervals and, on the right, the \gls{AUC} against the
  exact counterfactual, by which the rows are ordered. The two
  orderings disagree. Re-exposure cohorts (E) are near-duplicates of trained
  text; F, compared across surface forms, fails its placebo test; X is synthetic.}
  \label{fig:cohort}
\end{figure}

\paragraph*{Effect is uniform; detectability is not}
Across the real-text single-exposure cohorts, the mean
per-token effect lies in a relatively narrow range
(\num{\cohwgaptau} to \qty{\cohltau}{nats/token}),
yet \gls{AUC} varies from \num{\cohlauc} to
\num{\cohsseauc} (\Cref{fig:cohort,tab:cohorts}).
StackExchange posts, which are short and often formulaic,
reach an \gls{AUC} of \num{\cohsseauc} and
\qty{\cohssetpr}{\percent} \gls{TPR} at
\qty{1}{\percent} \gls{FPR}. In contrast, arXiv papers
have a comparable average effect (\num{\cohmtau}) but
reach only \num{\cohmauc} \gls{AUC}.
Thus, average effect size alone does not explain
detectability; variation in document scores also matters.
The standardized ratio partly addresses this variation
by scaling each document's reference ratio by its
estimated token-level noise. Its gains at
\qty{1}{\percent} \gls{FPR} are particularly large
for the short Wikipedia cohorts: from
\qty{\cohwnewtprorc}{\percent} to
\qty{\cohwnewtprzorc}{\percent} for W\textsubscript{new},
and from \qty{\cohwgaptprorc}{\percent} to
\qty{\cohwgaptprzorc}{\percent} for
W\textsubscript{gap}. For arXiv papers, the improvement
is smaller, from \qty{\cohmtprorc}{\percent} to
\qty{\cohmtprzorc}{\percent}
(\Cref{sec:results-signal}).
However, we cannot fully separate intrinsic document
variability from the uneven domain shifts caused by the
joint intervention and possible execution differences.
ArXiv received \qty{90}{\percent} of the injected
tokens and is among the least detectable cohorts,
so these shifts may contribute to its poorer detection
performance. Previously exposed text behaves differently:
re-exposing original Wikipedia content (E0) reaches
\num{\cohezeroauc} \gls{AUC}, while edited versions (E)
reach \num{\coheauc}. These high values show that
additional exposure to previously seen or closely
related text can produce stronger membership evidence
than a first exposure to unseen documents.

\paragraph*{Two cohorts fail}
Cohort F injects \LaTeX-derived text while its held-out
partner is evaluated in PDF-derived form. The control
model alone distinguishes the two arms with an
\gls{AUC} of \num{0.84}, showing that their surface
forms are confounded with membership. Its original
\gls{AUC} of $\cohfauc$ therefore cannot be interpreted
as membership detection.
We also rescore both arms in the same representation,
using \num{\cohfnform} pairs that pass the data audit
and have complete scores for both forms. The control
model then gives \gls{AUC} \num{\cohfplacebo} for
\LaTeX{} and \num{\cohfplacebopdf} for PDF text,
consistent with the removal of the original confound.
The within-pair effect measured on PDF renderings is
$\cohfrho$ (\gls{CI}
\numrange[range-phrase=--]{\cohfrholo}{\cohfrhohi})
of that measured on the \LaTeX{} sources. This provides
an exploratory estimate of how the effect of training
transfers across document representations.
Cohort X also requires separate treatment. Its synthetic
anchors show a relatively large mean effect
(\num{\cohxtau}) but only \num{\cohxauc} \gls{AUC}.
Scores for random-token and model-sampled text are
unstable across independent runs, producing extreme
values that would dominate low-\gls{FPR} thresholds.
We therefore exclude both F and X from pooled
negative sets.

\paragraph*{Longer documents help where token noise dominates}
The estimated token-sampling component accounts for
\qty{\tokshare}{\percent} of the within-pair variance
across verified-unseen single-exposure pairs.
For arXiv papers, it accounts for only
\qty{\toksharearxiv}{\percent}, indicating that
variation across documents is the dominant source of
uncertainty. For the shorter Wikipedia and StackExchange
documents, the share rises to
\qty{\tokshareshort}{\percent}, suggesting that
additional tokens could reduce a larger fraction of
their score variability.
To measure the benefit of additional text, we consider
the \num{\lennfixed} pairs in cohort M for which both
papers contain at least \num{4096} tokens. Using only
the first \num{16} tokens gives an \gls{AUC} of
\num{\lenaucsixteen}, which rises to $\lenaucfourk$
when scoring \num{4096} tokens (paired difference
\num{\lendiff}, \gls{CI}
\numrange[range-phrase=--]{\lendifflo}{\lendiffhi}).
Scoring the remainder of each paper adds almost
nothing (\num{\lendifffull}, \gls{CI}
\numrange[range-phrase=--]{\lendifffulllo}{\lendifffullhi}).
The truncated-paper cohort L shows a similar, though
less precise, pattern. Its \gls{AUC} rises from
\num{\cohlaucfivetwelve} at \num{512} tokens to
\num{\cohlaucfourk} at \num{4096} tokens, with only
\num{50} pairs per length and \qty{95}{\percent}
intervals of approximately $\pm\num{0.1}$.
Pooled over all lengths, its performance is comparable
to that of full papers (\num{\cohlauc} versus
\num{\cohmauc}). These results suggest that additional
tokens improve discrimination until token-level noise
becomes small relative to variation across documents.

\section{Discussion}
\label{sec:discussion}

\paragraph*{Two questions, one name}
The clearest implication of our measurements is that \enquote{membership inference}
refers to two questions whose evidence requirements differ by more than an order of
magnitude. With the exact counterfactual, \gls{AUC} is \num{\aucnat} and \qty{\tpronenat}{\percent}
of single-exposure documents are caught at \qty{1}{\percent} \gls{FPR} with a pooled threshold
(\qty{\tpronezorcnat}{\percent} with the standardized ratio), \qty{\tprstratorc}{\percent}
(\qty{\tprstratzorc}{\percent}) with one threshold per domain; the same scores identify a
collection of \num{\npower} matched pairs with \qty{80}{\percent} power. Individual membership inference is more effective for short
StackExchange posts when domain-matched non-members are
available. For long documents, the evidence is much weaker.
For example, when half the candidate pool consists of members,
the standardized ratio identifies only \num{\screenmzfiftytrue}
of \num{\cohmn} arXiv papers at an \gls{FDR} level of $0.05$,
while the plain ratio identifies about \num{4}.
When the member prevalence drops to \qty{5}{\percent}
or below, neither identifies any arXiv members.
For the rest, the signal is present, but
the error rate at any useful threshold is unacceptable for an individual
accusation. A claim that a corpus was used is a different
matter and is well supported by a few dozen documents. Duplication helps a
claimant substantially --- works appearing four to eight times cross into
testable territory --- and post-training is not a defense: after our re-run of
the official recipe, the model retains
\qty{\ptinstructretain}{\percent} of the effect; forgetting, on the other hand, is
real, although under this learning-rate schedule, a document seen early ends the stage no
fainter than one seen late. We would therefore propose that papers making
document-level claims report the \gls{TPR} at a low \gls{FPR} and the precision at a stated prevalence, and that papers reporting only
\gls{AUC} describe their result as collection-level. Both document-level
measures come from prior work, the first from \citet{carlini2022membership}
and the second from \citet{jayaraman2021revisiting}; our contribution is to put a number on the gap between the two kinds of claim in a
setting where the ground truth is known by construction.

\paragraph*{What the collection-level result requires}
Our collection-level results depend on two important conditions.
First, members and non-members must come from comparable populations.
The domain-level shifts observed in \Cref{sec:results-spillover}
can otherwise produce misleading results, and these systematic
differences do not disappear when more documents are tested.
For example, a collection of \num{\npower} never-injected arXiv papers
is incorrectly identified as members in more than
\qty{99.9}{\percent} of trials when compared against
StackExchange non-members. This shows how differences between
domains can be mistaken for evidence of membership.
Second, the reference model must be sufficiently close to the
counterfactual. With the exact counterfactual, our paired test
requires only \num{\npower} matched pairs to achieve
\qty{80}{\percent} power. Using the published checkpoint instead
increases this requirement to \num{\tierpubnpower} pairs.
These findings are also relevant to dataset-inference
methods~\cite{maini2024llm,puerto2025scaling}, which compare
suspected training documents against held-out data.
For example, \citet{maini2024llm} assume that both collections
come from the same distribution. Our results reinforce the
importance of this assumption and show how reference quality
affects the amount of evidence required. However, the collection
sizes measured in our controlled experiment should not be
interpreted as general requirements for real-world membership
audits.

\paragraph*{What the negative results do not show}
Our reference-free attacks perform close to chance against
difficulty-matched non-members. This does not mean that
membership evidence is absent: our experiment measures a
training effect that these statistics largely fail to recover.
Conversely, success on unmatched populations may reflect
document difficulty or domain differences rather than membership.
Our findings concern mainly single-exposure documents and
attacks based on token log-probabilities. The standardized
ratio improves detection at low \gls{FPR}, but most long
documents remain difficult to identify. Under our
single-reference, shared-variance approximation, offline
LiRA~\cite{carlini2022membership} reduces to the reference
ratio. We do not evaluate attacks using gradients,
activations, or additional model queries. Repeated exposure
can also substantially improve detection (\Cref{fig:dose}).
Finally, the exact counterfactual provides an exceptionally
strong reference that real-world adversaries are unlikely
to obtain. It represents a near-best-case reference for
the evaluated attacks, although it is not a theoretical
upper bound on their performance. The choice of statistic
still matters: combining the reference ratio with Min-K\%
improves \gls{AUC} over the plain ratio
(\Cref{sec:results-attacks}). Our negative results therefore
concern the evaluated attacks and conditions, not membership
inference in general.

\paragraph*{Limitations}
We study one 1B-parameter model during its 50-billion-token
mid-training stage and subsequent post-training. Re-running
the preceding 4-trillion-token pretraining stage was beyond
our computational budget. Further studies are needed to
determine whether our findings hold for larger models,
longer training runs, and different learning-rate schedules.
We inject \num{\npairsdesign} documents jointly, replacing
\qty{0.135}{\percent} of the training stream. Our estimates
therefore concern this joint intervention, not the isolated
effect of including one document. Injected documents begin
at training-window boundaries, which may favor detectability.
Our verification detects verbatim overlaps of at least
40 words, but shorter or heavily paraphrased overlaps may
remain. The retrained reference shares the official data
mixture and may be stronger than one an adversary could
reproduce. Precision depends on member prevalence, while
our \gls{FDR} results require appropriate calibration
non-members. Several analyses, including the standardized
ratio, domain-specific thresholds, collection-level tests,
training trajectories, and post-training comparison, were
exploratory. Finally, cohort F's original comparison is
confounded by document representation, and a few injections
were affected by training-data filtering or displacement
(\Cref{app:cohorts,app:implementation}).
\section{Related Work}
\label{sec:related}
\paragraph*{Attacks, benchmarks, and critiques}
Loss-threshold attacks~\cite{yeom2018privacy} and calibrated variants --- zlib~\cite{carlini2021extracting}, Min-K\%~\cite{shi2024detecting}, Min-K\%++~\cite{zhang2025min}, ReCaLL~\cite{xie2024recall}, and reference models~\cite{sablayrolles2019white,watson2022importance,ye2022enhanced,carlini2022membership,hayes2025exploring} --- define standard practice. However, observational evaluations frequently confound membership with distribution shift: blind classifiers match reported attacks on uncurated splits~\cite{das2025blind}, and same-distribution benchmarks show performance dropping to near chance~\cite{duan2024membership,meeus2025sok,shi2026olmo}. While recent datasets leverage open corpora (\hbox{CheckMIABench}~\cite{wang2026checkmiabench}, OLMoMIA~\cite{kim2026detecting}, OLMo-Detect~\cite{shi2026olmo}), they lack counterfactual controls. Our testbed provides the randomized, same-data baseline these critiques demand.

\paragraph*{Controlled memorization and dynamics} Prior controlled studies rely on synthetic canaries~\cite{carlini2019secret}, watermarks~\cite{wei2024proving,sander2024watermarking}, test contamination~\cite{bordt2025much,wei2026correcting}, or continued-pretraining insertions~\cite{huang2024demystifying,chang2024large}. Closest to our work, \citet{wei2025hubble} insert passages into models trained from scratch, and \citet{jagielski2023measuring} track example forgetting. We extend this to real documents injected across a complete official mid-training stage with an exact same-seed twin, tracking \num{24} step-matched checkpoints and the final models through post-training. This characterizes the interaction between decay and learning-rate annealing~\cite{hgele2024scaling,tissue2025scaling}, quantifies domain spillover~\cite{grosse2023studying,shilov2026mosaic}, and establishes that while post-training retains \qty{\ptinstructretain}{\percent} of the exposure signal~\cite{nasr2023scalable,zhang2024persistent}, collection-level audits~\cite{maini2024llm,puerto2025scaling} remain far more statistically tractable than individual document attribution.

\paragraph*{Post-training}
Prior studies show that aligned chat models can emit memorized pretraining data~\cite{nasr2023scalable}, data-poisoning effects persist through alignment~\cite{zhang2024persistent}, and fine-tuning memorization survives \gls{RLHF}~\cite{hayes2025measuring}. We evaluate what the official post-training recipe does to membership traces against a byte-matched counterfactual: while \qty{\ptinstructretain}{\percent} of the raw effect survives, the signal is not sharpened for single-exposure documents.

\paragraph*{Document-level and collection-level inference}
Document-level attacks typically train classifiers on summary features of normalized token probabilities~\cite{meeus2024did}, whereas collection-level methods test datasets against held-out splits~\cite{maini2024llm,puerto2025scaling} or synthetic distributions~\cite{zhao2025unlocking}, or leverage exchangeability tests for contamination~\cite{oren2024proving}. Our evaluation quantifies the orders-of-magnitude separation between these two objectives within a single controlled experiment: the published-checkpoint tier demonstrates why collection-level audits succeed with moderate sample sizes even when single-document attribution remains statistically intractable.

\section{Conclusion}
\label{sec:conclusion}

We constructed a randomized, closely matched counterfactual
for the complete 50-billion-token mid-training stage of
OLMo~2~1B. Our experiment includes \num{\npairsdesign}
matched document pairs, nearly all real text, with one
document per pair randomly selected for injection. By
comparing the treated model against \gls{AI2}'s released
same-seed run and tracking \num{24} step-matched checkpoints
across three training ladders, we measure how exposure
changes document likelihood and whether that change
supports membership inference.

Our central finding is that a measurable training effect
does not necessarily support reliable individual membership
decisions. A single exposure produces an average within-pair
effect of \qty{\taunat}{nats/token}, with
\qty{\posnat}{\percent} of pairs correctly ordered and
an observed mean \num{\permz} \glspl{sd} from the
randomization null. Yet even with the exact counterfactual,
the reference ratio achieves only
\qty{\tpronenat}{\percent} \gls{TPR} at
\qty{1}{\percent} \gls{FPR} using a pooled threshold.
Token-level standardization improves this to
\qty{\tpronezorcnat}{\percent}, while cohort-specific
calibration raises detection to
\qty{\tprstratorc}{\percent} and
\qty{\tprstratzorc}{\percent}, respectively, given
suitable non-members for each cohort. Under
false-discovery control at $q=0.05$ and
\qty{50}{\percent} member prevalence, the standardized
ratio identifies about \num{\screenzfiftytrue} of
\num{\npairsnat} members, compared with none for the
plain ratio. At \qty{5}{\percent} prevalence or below,
it identifies essentially none.

In contrast, aggregating evidence across matched documents
is substantially more effective. With the exact
counterfactual, a paired test requires an estimated
\num{\npower} matched pairs for \qty{80}{\percent}
power, assuming comparable non-members. An independently
seeded reference also recovers \qty{\sibshare}{\percent}
of the counterfactual's \gls{AUC} advantage over chance.
Thus, the difficulty of individual attribution cannot
be explained by reference quality alone; substantial
variation in document-level effects remains even with
a closely matched reference.

The remaining experiments help explain this gap.
Reference-free attacks recover little additional evidence
once document difficulty is matched, while documents never
injected into training also exhibit domain-dependent
likelihood shifts. Repeated exposure increases the average
effect nearly linearly through \num{32} copies, whereas
checkpoint trajectories reveal substantial decay after
individual exposures that is difficult to detect from
the final checkpoint alone. The official post-training
recipe preserves approximately
\qty{\ptinstructretain}{\percent} of the average effect,
although individual detectability varies considerably
across document types. Together, these findings show
why measurable exposure effects, individual attribution,
and collection-level inference must be evaluated
separately.

Our results concern one model size, one mid-training
recipe, and the evaluated attack statistics. The testbed
provides a basis for studying whether these findings
change with model scale and learning-rate schedules.
We will release the treated checkpoints, post-trained
models, per-token log-probabilities, and frozen assignment
manifest, enabling future work to measure membership
evidence against the same controlled reference.
This work can specifically help researchers design their
attacks using our findings, and the checkpoints and cohorts provide
them with a domain- and difficulty-matched testbed to assess their methods.
\section*{Acknowledgments}

This work was supported by the Knut and Alice Wallenberg Foundation through a Wallenberg
Scholar Grant to Prof. Dejan Kosti\'c and through computational resources on the Berzelius supercomputer at the National Academic Infrastructure for Supercomputing in Sweden.

\clearpage  %

\printbibliography

\appendices

\section{The Thirteen Cohorts}
\label[appendix]{app:cohorts}
\Cref{tab:cohorts-design} lists every cohort as built. All documents are real
text except cohort X. The E cohorts are, by design, edited versions of
Wikipedia pages present in both training stages, so their held-out partners
are members with a known trained ancestor; they measure re-exposure, not
membership, and are excluded from the verified-unseen population. A \foreignlanguage{latin}{post hoc}
audit found that the official chunks displaced by injected copies contained
the ancestors of some E-family documents, so the control saw text that the
treated run did not: \num{49} evaluation documents share at least \num{50}
document-specific 13-grams with displaced text, \num{17} injected and \num{13} held-out
among the pairs, with no consistent sign, so this adds noise to the
re-exposure cohorts rather than bias.

\begin{table}[th]
  \centering
  \caption{\textbf{Cohorts as built.} Pairs, copies per document $r$ and
  tokens injected (rows rounded to \num{0.1}M; the total is exact). The pooled analyzed set has \num{\npairspooled} pairs after exclusions.}
  \label{tab:cohorts-design}
  \footnotesize
  \setlength{\tabcolsep}{3.5pt}
  \begin{tabular}{llrrr}
  \toprule
  Cohort & Source & Pairs & $r$ & Tokens \\
  \midrule
  M & arXiv, 11 months, 5 categories & \num{1986} & 1 & 17.2M \\
  L & arXiv prefixes, \qtyrange[range-units=single,range-phrase=--]{512}{8192}{tokens} & \num{250} & 1 & 0.8M \\
  W\textsubscript{new} & Wikipedia, created after cutoff & \num{1000} & 1 & 0.8M \\
  W\textsubscript{gap} & Wikipedia, absent from snapshot & \num{600} & 1 & 0.8M \\
  S\textsubscript{se} & StackExchange, 5 sites & \num{1000} & 1 & 0.8M \\
  D & arXiv, $r\in\{2,4,8,16\}$ & \num{400} & 2--16 & 25.8M \\
  D32 & arXiv & \num{40} & 32 & 10.8M \\
  DB & arXiv, 4 copies in \qty{1}{\percent} window & \num{60} & 4 & 2.1M \\
  F & arXiv, \LaTeX\ vs.\ PDF form & \num{278} & 1 & 4.2M \\
  X & synthetic anchors & \num{200} & 1 & 0.2M \\
  E0 & edited Wikipedia, \qty{0}{\percent} novel & \num{300} & 1 & 0.3M \\
  E0-D & edited Wikipedia, \qty{0}{\percent} novel & \num{200} & 4 & 0.8M \\
  E & edited Wikipedia, 5 strata & \num{1500} & 1 & 2.9M \\
  \midrule
  Total & & \num{7814} & & 67.3M \\
  \bottomrule
  \end{tabular}
\end{table}

\section{Attack Statistics}
\label[appendix]{app:attacks}
For a document of scored tokens $t=1,\dots,T$ with log-probabilities $\ell_t$
under the target model, and with $\mu_t$, $\sigma_t$ the mean and standard
deviation of the log-probability under the model's next-token distribution at
position $t$:
loss is $\frac1T\sum_t\ell_t$; perplexity is $\exp(-\mathrm{loss})$, and
bits-per-byte is $-\sum_t\ell_t/(\ln 2\cdot\text{bytes})$; both are negated so
that larger is more member-like; the zlib ratio is $\sum_t\ell_t$ divided by
the zlib-compressed length of the text, a variant of the ratio of
\citet{carlini2021extracting} with the summed rather than the mean
log-probability;
\hbox{Min-K\%}~\cite{shi2024detecting} is the mean of the $K\%$ smallest $\ell_t$;
\hbox{Min-K\%++}~\cite{zhang2025min} is the mean of the $K\%$ smallest
$(\ell_t-\mu_t)/\sigma_t$; and surprise is $\frac1T\sum_t(\ell_t-\mu_t)$. A
reference-based variant subtracts the same statistic under the reference
model. We evaluated $K\in\{5,10,20,30,40,50\}$ and report $K{=}20$; choosing
$K$ on half of the pairs and evaluating on the other half changes the pooled
\glspl{AUC} by less than \num{0.001}.

\paragraph*{Standardized ratio} With $d_t=\ell_t-\ell^{\mathrm{ref}}_t$, the
per-token difference against the reference and $T$ the number of scored
tokens, the statistic is $\bar d\,/\,(s_d/\sqrt{T_{\mathrm{eff}}})$, where
$s_d$ is the sample \gls{sd} of $d_t$ and
$T_{\mathrm{eff}}=T/\tau$, with $\tau=1+2\sum_{k\le W}\rho_k$ the integrated
autocorrelation of $d_t$, the window $W$ the smallest $W\ge5\,\tau(W)$
\cite{sokal1997monte}, and $T_{\mathrm{eff}}$ clamped to $[1,T]$; the median
$\tau$ against the exact counterfactual is \num{\calibtaumedian}, so the median
document keeps \qty{\calibneffshare}{\percent} of its tokens. Positions without a score,
the first of each scoring window, are dropped. Using $T$ in place of
$T_{\mathrm{eff}}$ gives \gls{AUC} \num{\calibzTorcauc} against
\num{\calibzorcauc} and a per-domain rate of \qty{\calibzTorcstrat}{\percent} against
\qty{\calibzorcstrat}{\percent}.

\begin{table}[t]
  \centering
  \caption{\textbf{The membership signal across post-training} on
  verified-unseen single-exposure pairs. Stages are cumulative. RLVR1 is the end
  of the first \gls{RLVR} stage; RLVR2 (2k) is step~2000 of the second stage, the
  checkpoint corresponding to the released OLMo~2~1B~Instruct, and RLVR2 (end)
  is that stage's final step (2604).}
  \label{tab:stages}
  \footnotesize
  \setlength{\tabcolsep}{3.5pt}
  \begin{tabular}{@{}l S[table-format=+1.4] S[table-format=1.4] S[table-format=1.2] S[table-format=1.3] S[table-format=1.1]@{}}
\toprule
Stage & {$\hat\tau$} & {SD} & {$d_z$} & {\auc} & {\tprat{1} (\%)} \\
\midrule
Mid-training & +0.0221 & 0.0309 & 0.71 & 0.767 & 3.8 \\
+ SFT & +0.0155 & 0.0239 & 0.65 & 0.731 & 3.7 \\
+ SFT + DPO & +0.0199 & 0.0343 & 0.58 & 0.708 & 4.8 \\
+ RLVR1 & +0.0196 & 0.0350 & 0.56 & 0.700 & 5.1 \\
+ RLVR2 (2k) & +0.0196 & 0.0355 & 0.55 & 0.697 & 4.7 \\
+ RLVR2 (end) & +0.0197 & 0.0355 & 0.55 & 0.699 & 4.9 \\
\bottomrule
\end{tabular}

\end{table}

\begin{table}[t]
  \centering
  \caption{\textbf{Attack \gls{AUC} and true-positive rate at \qty{1}{\percent} \gls{FPR}}, grouped by
  the reference the adversary holds; verified-unseen single-exposure pairs
  against their matched partners.}
  \label{tab:attacks}
  \footnotesize
  \setlength{\tabcolsep}{4pt}
  \begin{tabular}{l S[table-format=1.3] S[table-format=1.1]}
\toprule
Attack & {\auc} & {\tprat{1} (\%)} \\
\midrule
\multicolumn{3}{l}{\emph{Exact counterfactual}} \\
\quad Ratio + Min-K\% & 0.816 & 5.0\\
\quad Standardized ratio & 0.778 & 8.8\\
\quad Ratio & 0.767 & 3.8\\
\quad Ratio + Min-K\%++ & 0.665 & 1.6\\
\addlinespace
\multicolumn{3}{l}{\emph{Retrained reference}} \\
\quad Ratio + Min-K\% & 0.779 & 5.3\\
\quad Ratio & 0.745 & 3.7\\
\quad Standardized ratio & 0.725 & 8.1\\
\quad Ratio + Min-K\%++ & 0.641 & 1.2\\
\addlinespace
\multicolumn{3}{l}{\emph{Published checkpoint}} \\
\quad Ratio + Min-K\% & 0.626 & 1.6\\
\quad Ratio + Min-K\%++ & 0.608 & 1.6\\
\quad Ratio & 0.603 & 1.3\\
\quad Standardized ratio & 0.562 & 2.0\\
\addlinespace
\multicolumn{3}{l}{\emph{Reference-free}} \\
\quad Surprise & 0.566 & 1.6\\
\quad Min-K\%++ ($K{=}20$) & 0.559 & 1.5\\
\quad Min-K\% ($K{=}20$) & 0.527 & 1.3\\
\quad Bits-per-byte & 0.519 & 1.2\\
\quad LOSS & 0.519 & 1.4\\
\quad Perplexity & 0.519 & 1.4\\
\quad zlib ratio & 0.517 & 1.2\\
\bottomrule
\end{tabular}

\end{table}

\begin{table}[t]
  \centering
  \caption{\textbf{The reference ratio and the standardized ratio.} Every row is
  evaluated on the same \num{\npairsnat} verified-unseen single-exposure pairs,
  with pair-bootstrap \qty{95}{\percent} intervals; \enquote{per cohort} is the
  \gls{TPR} at \qty{1}{\percent} \gls{FPR} with one threshold per cohort.}
  \label{tab:calibrated}
  \scriptsize
  \setlength{\tabcolsep}{3pt}
  \begin{tabular}{@{}l S[table-format=1.3]@{\,}l S[table-format=1.1]@{\,}l S[table-format=2.1]@{}}
\toprule
Attack & \multicolumn{2}{c}{\auc\ [95\% CI]} & \multicolumn{2}{c}{\tprat{1} [95\% CI]} & {per cohort} \\
\midrule
\multicolumn{6}{l}{\emph{Exact counterfactual}} \\
\quad Ratio & 0.767& {\scriptsize[.759,.774]} & 3.8& {\scriptsize[2.9,4.8]} & 16.5 \\
\quad Standardized ratio & 0.778& {\scriptsize[.771,.783]} & 8.8& {\scriptsize[7.3,10.9]} & 31.4 \\
\addlinespace
\multicolumn{6}{l}{\emph{Retrained reference}} \\
\quad Ratio & 0.745& {\scriptsize[.737,.753]} & 3.7& {\scriptsize[2.8,4.8]} & 8.2 \\
\quad Standardized ratio & 0.725& {\scriptsize[.718,.731]} & 8.1& {\scriptsize[6.0,9.5]} & 20.2 \\
\addlinespace
\multicolumn{6}{l}{\emph{Published checkpoint}} \\
\quad Ratio & 0.603& {\scriptsize[.593,.613]} & 1.3& {\scriptsize[0.8,1.9]} & 2.8 \\
\quad Standardized ratio & 0.562& {\scriptsize[.557,.566]} & 2.0& {\scriptsize[1.6,2.9]} & 2.6 \\
\bottomrule
\end{tabular}

\end{table}

\begin{table}[t]
  \centering
  \caption{\textbf{Per-cohort effect and detectability} against the exact
  counterfactual, with the \qty{1}{\percent} \gls{FPR} threshold set within each cohort; $r$ is
  copies per document and $n$ is pairs after exclusions.}
  \label{tab:cohorts}
  \footnotesize
  \setlength{\tabcolsep}{3.5pt}
  \setlength{\tabcolsep}{3pt}
\begin{tabular}{@{}ll S[table-format=4.0] c S[table-format=+1.4] S[table-format=1.3] S[table-format=3.1]@{}}
\toprule
Cohort & Note & {$n$} & $r$ & {$\hat\tau$} & {\auc} & {\textsc{tpr} (\%)} \\
\midrule
D32 &  & 40 & 32 & +0.7258 & 1.000 & 100.0 \\
S\textsubscript{se} &  & 1000 & 1 & +0.0251 & 0.962 & 52.0 \\
D &  & 400 & 2--16 & +0.1508 & 0.943 & 56.2 \\
DB &  & 60 & 4 & +0.0757 & 0.921 & 21.7 \\
W\textsubscript{gap} &  & 600 & 1 & +0.0178 & 0.900 & 12.7 \\
W\textsubscript{new} &  & 1000 & 1 & +0.0182 & 0.874 & 8.5 \\
M &  & 1982 & 1 & +0.0232 & 0.718 & 5.5 \\
L &  & 250 & 1 & +0.0264 & 0.713 & 2.8 \\
E0-D & re-exposure & 200 & 4 & +0.0669 & 0.997 & 100.0 \\
E0 & re-exposure & 299 & 1 & +0.0164 & 0.957 & 36.5 \\
E & re-exposure & 1495 & 1 & +0.0179 & 0.950 & 17.7 \\
X & synthetic & 200 & 1 & +0.0564 & 0.645 & 12.5 \\
F & failed placebo & 277 & 1 & +0.0094 & 0.627 & 1.1 \\
\bottomrule
\end{tabular}

\end{table}

\section{Statistical Details}
\label[appendix]{app:stats}
\paragraph*{Randomization test} Under the null that injection has no effect,
each within-pair contrast is equally likely to carry either sign, so the null
distribution of the mean is obtained by flipping signs at random. Its standard
deviation is $\sqrt{\sum_i d_i^2}/n$, which gives the standard score reported
in the text; a Monte Carlo draw of $10^5$ sign vectors never reached the
observed mean.

\paragraph*{Operating points} The threshold for an \gls{FPR} $f$ is
the score that admits $\lfloor f\cdot n_{\mathrm{neg}}\rfloor$ held-out
partners above it; with $n_{\mathrm{neg}}=\negnat$ this is the \num{49}\textsuperscript{th} largest
held-out score at \qty{1}{\percent} and the \num{5}\textsuperscript{th} at \qty{0.1}{\percent}. For cohorts with fewer than
\num{100} negatives the \qty{1}{\percent} threshold is the single largest held-out score (the second largest with \numrange{100}{199} negatives), and
the rate is at the design's resolution floor.

\paragraph*{Naming documents under false-discovery control} The pairs are
split into five folds, balanced within cohorts. For each fold, a candidate
pool combines the fold's non-members with a random subset of its members
chosen to give the stated prevalence. A candidate with score $s$ receives the
conformal $p$-value $(1+\#\{\text{calibration non-members scoring at least }s\})/(1+n_{\mathrm{cal}})$,
computed against the \num{\screenncal} non-members of the other four folds, and
the Benjamini--Hochberg procedure at level $q$ is applied within the pool.
Counts are totals over the five folds, averaged over \num{50} random fold
assignments; ranges are the \num{2.5}\textsuperscript{th} and \num{97.5}\textsuperscript{th} percentiles over
assignments. Because the smallest attainable $p$-value is
$1/(1+n_{\mathrm{cal}})$, a pool that is half members can name a document
only if at least \num{\screenzfiftyminrej} of its candidates score above every
calibration non-member.

\paragraph*{Collection-level AUC} For each $m$ we draw $m$ members and $m$
non-members independently with replacement from the population, compare the
two means, and repeat at least \num{400} times; \gls{AUC} is the rank \gls{AUC} of the
resulting pooled scores. The draws are unpaired, so the matched partners are
not used as the non-member collection.

\paragraph*{Token-noise share} From the per-token contrasts we estimate for
each document the variance of its mean effect under token resampling, using
the effective number of independent tokens from a windowed autocorrelation
estimate, and report the mean over pairs of the two partners' token variances
divided by the variance of the within-pair contrast.

\paragraph*{Rescoring}
Within each comparison of a treated model with its control, both are scored
with the same code. Mid-training checkpoints are scored with OLMo's model code
and post-trained models with the Hugging Face implementation of the same
architecture (\Cref{app:implementation}). In fp32 on CPU the two give bitwise
identical log-probabilities on identical weights; in bf16 they differ only by
rounding: on \num{1121} documents scored on GPU under the stage-1 endpoint
with both, per-document means differ by a median of $2.3\times10^{-4}$
nats/token and at most $2.9\times10^{-3}$, with mean $1\times10^{-5}$
(standard error $1.6\times10^{-5}$). The one difference between arms is the
attention kernel at mid-training: the treated checkpoints were saved with flash
attention enabled, whereas \gls{AI2}'s checkpoints use PyTorch's
scaled-dot-product attention, as do all post-trained models. Flash attention
has no fp32 implementation, so we rescored \num{327} documents of at most
\num{1400} tokens under both mid-training endpoints in fp32 on CPU with one
implementation and compared the stored bf16 scores with this reference. The
error of the per-document effect averages $-3\times10^{-5}$ nats/token
(standard error $7\times10^{-5}$, per-document \gls{sd} $1.2\times10^{-3}$);
its difference between injected and held-out documents, $-1.4\times10^{-4}$
(standard error $1.6\times10^{-4}$), is at most \qty{2}{\percent} of the
within-pair effect at \qty{95}{\percent} confidence, and on never-injected
documents it averages $-3\times10^{-4}$.

\section{Implementation Details}
\label[appendix]{app:implementation}

\begin{table*}[t]
  \centering
  \caption{\textbf{Training settings.} Our mid-training run follows \gls{AI2}'s
  published stage-2 configuration; post-training settings are the same for
  the treated model and the control. Dataset revisions are the first eight
  hexadecimal digits of the Hugging Face commit.}
  \label{tab:hyperparameters}
  \small
  \setlength{\tabcolsep}{4pt}
  \renewcommand{\arraystretch}{1.1}
  \begin{tabular}{@{}>{\raggedright\arraybackslash}p{1.7cm}
                    >{\raggedright\arraybackslash}p{3.6cm}
                    >{\raggedright\arraybackslash}p{3.35cm}
                    >{\raggedright\arraybackslash}p{3.35cm}
                    >{\raggedright\arraybackslash}p{4.05cm}@{}}
  \toprule
  & Mid-training & \gls{SFT} & \gls{DPO} & \gls{RLVR} (each stage) \\
  \midrule
  Start & stage-1 endpoint, with optimizer state & mid-training endpoint &
    final \gls{SFT} model & final \gls{DPO} model; final RLVR1 model \\
  Data & \num{576} stage-2 files, seed-42 order &
    allenai/\allowbreak tulu-3-sft-olmo-2-mixture-0225 @\,\texttt{d91a0785} &
    allenai/\allowbreak olmo-2-0425-1b-preference-mix @\,\texttt{c6454f3d} &
    allenai/\allowbreak RLVR-GSM-MATH-IF-Mixed-Constraints @\,\texttt{7dbd180f};
    allenai/\allowbreak RLVR-MATH @\,\texttt{bd2a9355} \\
  Size & \num{12212224} of \num{12570760} chunks & \num{861812} examples &
    \num{371767} pairs & \num{29689}; \num{7499} prompts \\
  Max.\ length & \num{4096} & \num{4096} & \num{2048} &
    \num{2048} prompt, \num{2048} response \\
  Objective & cross-entropy, z-loss $10^{-4}$ (configured $10^{-5}$) &
    cross-entropy on assistant tokens, mean per micro-batch &
    length-normalized \gls{DPO}, $\beta=5$ &
    policy gradient with group-normalized advantages, Kullback--Leibler
    coefficient \num{0.01}; reward $10\times$ verifier score \\
  Learning rate & \num{7.4487e-5}, linear to 0, no warm-up &
    \num{3e-5}, linear to 0, \qty{3}{\percent} warm-up &
    \num{2.5e-6}, linear to 0, \qty{10}{\percent} warm-up & \num{5e-7}, constant \\
  Optimizer & AdamW, $\beta_{1,2}=(0.9, 0.95)$, $\epsilon=10^{-8}$, weight
    decay \num{0.1} (not on embeddings) &
    AdamW, $\beta_{1,2}=(0.9, 0.999)$, $\epsilon=10^{-8}$, weight decay 0 &
    as \gls{SFT} & as \gls{SFT}, but weight decay \num{0.01} \\
  Batch & $512\times\num{4096}$ tokens & 128 sequences & 128 pairs &
    48 prompts $\times$ 16 responses, 2 updates per step \\
  Duration & \num{23852} updates & 2 epochs, \num{13466} updates &
    1 epoch, \num{2905} updates & \num{2604} steps \\
  Gradient clip & 1.0 & 1.0 & 1.0 & 1.0 \\
  Precision & bf16, fp32 gradient reduction & bf16, ZeRO-3 &
    bf16, ZeRO-2 & bf16, ZeRO-2 \\
  Seed & 42 & 1 & 111 & 1 \\
  GPUs & $2\times8$ A100-SXM4-80GB & $8$ A100-SXM4-80GB & $8$ A100-SXM4-80GB &
    $2\times8$ A100-SXM4 (12 for training, 4 for sampling) \\
  Checkpoints & every \num{1000} updates & every \num{1000} updates &
    every 500 updates & every 200 steps \\
  \bottomrule
  \end{tabular}
\end{table*}

This appendix gives the implementation details of the training runs and the
analysis; \Cref{tab:hyperparameters} lists the training settings. Our scripts
and configuration files are part of the code we will release.

\paragraph*{Mid-training run}
The run starts from \gls{AI2}'s OLMo~2~1B stage-1 endpoint at step
\num{1907359}, which is the released pre-stage checkpoint of
\Cref{sec:design-twin} and the published checkpoint of \Cref{sec:background}.
We downloaded it as an OLMo-native checkpoint with model weights and AdamW
state from
\url{https://olmo-checkpoints.org/ai2-llm/peteish1/step1907359-unsharded/},
the load path of \gls{AI2}'s published stage-2 configuration. On the four
tensors we compared, the Hugging Face branch
\texttt{stage1-\allowbreak{}step1907359-\allowbreak{}tokens4001B} of
\texttt{allenai/\allowbreak{}OLMo-\allowbreak{}2-\allowbreak{}0425-\allowbreak{}1B}
is bitwise identical to it. Weights and optimizer state are loaded, and the
update counter and data position start at zero, with
\texttt{restore\_\allowbreak{}dataloader},
\texttt{reset\_\allowbreak{}optimizer\_\allowbreak{}state} and
\texttt{reset\_\allowbreak{}trainer\_\allowbreak{}state} set as in
\gls{AI2}'s seed-42 configuration; OLMo restored no
random-number-generator state from the checkpoint, and all dropout rates
are zero. The trainer is OLMo at commit
\texttt{090253da\allowbreak{}c6688f25\allowbreak{}32509daa\allowbreak{}7aa2eb5f\allowbreak{}ae50e956},
run in a container image built from
\texttt{nvcr.io/\allowbreak{}nvidia/\allowbreak{}pytorch:26.06-\allowbreak{}py3}
(PyTorch 2.13.0a0, CUDA 13.3, NCCL 2.30.5, flash-attn 2.7.4.post1, NumPy
1.26.4). The configuration is \gls{AI2}'s published
\texttt{configs/\allowbreak{}official-\allowbreak{}0425/\allowbreak{}OLMo2-\allowbreak{}1B-\allowbreak{}stage2-\allowbreak{}seed42.yaml}
at that commit, in which we changed only the data paths, run name, output
folder, checkpoint intervals and retention, and experiment-tracking keys.
Command-line options also set the load path to our local copy, disabled the
\num{27} in-loop downstream evaluations, enabled resumption from the latest
checkpoint and, from the third launch, turned off ephemeral checkpoints. The
file sets the z-loss multiplier
(\texttt{auxiliary\_\allowbreak{}loss\_\allowbreak{}multiplier}) to
$10^{-5}$, but the pinned trainer does not pass it to the loss function,
which applies its default of $10^{-4}$. The trainer derives the
learning-rate horizon from the duration of $5\times10^{10}$ tokens, so the
rate reaches zero at update \num{23842} and stays zero for the last \num{11}
of the \num{23852} updates; the per-update logs confirm this.

The stage-2 data are the \num{576} pre-tokenized files listed in the
configuration (\num{51491007838} tokens, or \num{12570760} chunks of
\num{4096} tokens when each file is cut separately); \gls{AI2}'s seed-42 and
seed-42069 configurations list the same files in the same order. OLMo's
loader permutes the chunk indices once with NumPy's PCG64 generator seeded
with the training seed (42), and update $u$ reads positions
$[512(u{-}1),512u)$ of this order for any number of GPUs, so the \num{23852}
updates consume its first \num{12212224} positions (\qty{97.1}{\percent} of
the chunks). We compute this order for the official files
and, separately, the order the same loader produces for a set of
\num{12212224} chunks; we then write the official chunk at each position $p$
of the first order, or the injected chunk that replaces it, into the slot that
the second order reads at position $p$, in \num{64} new files. With the data
paths pointed at these files, the unmodified loader reads at every position
the chunk of \gls{AI2}'s sequence or its substitute. Injected documents were
tokenized with the Hugging Face tokenizer of
\texttt{allenai/\allowbreak{}OLMo-\allowbreak{}2-\allowbreak{}0425-\allowbreak{}1B},
without special tokens and with one end-of-text token (id \num{100257})
appended; this tokenizer has the same vocabulary, merges and special tokens as
the trainer's. Of the copies placed as described in
\Cref{sec:design-injection}, \num{25} were moved from their targets, by at
most \num{359} positions; the first injected chunk falls in update \num{240}
and the last in update \num{23614}. OLMo's instance filter, set as in
\gls{AI2}'s configuration, masks the loss of any chunk in which a sequence of
\num{1} to \num{13} tokens repeats at least \num{32} times in a row; it masked
\num{43} injected chunks from \num{18} documents, including the only chunk of
two single-copy documents, which therefore received no gradient at all. We
keep these documents in the analyses.

Before training, the SHA-256 hashes of all \num{64} files were checked against
those recorded when the files were written, all \num{24081} injected chunks
were compared with the frozen token files of the injected documents, and
\num{20000} sampled background chunks with the official files. Before each of
the five launches, a check repeated the comparison for all injected chunks and
for the same \num{5000} background positions each time, compared file sizes,
and verified the SHA-256 hashes of the configuration, the file listing the
positions of the injected copies, the container image and the four stage-1
checkpoint files. After training, the
global index file that OLMo saved equals the second order above; the per-rank
data indices it logged for updates \num{18001}--\num{23852}, the only ones
kept, match it in all \num{93632} rows; read through this index file, all
\num{12414} injected copies hold their documents' tokens at their planned
positions; and \num{3700} non-injected positions, drawn at random in two
independent samples, are byte-identical to the official chunks at those
positions.

Each of the \num{16} GPUs processed its \num{32} sequences per update as eight
micro-batches of four. Training ran as five launches; four resumed from
the run's own checkpoints with data-loader and random-number-generator state
restored, and the retained checkpoints form one continuous run. The only
changes to the trainer concern writing checkpoints to the shared file system;
no loss, optimizer or data code was changed. The configuration stored in
\gls{AI2}'s seed-42 checkpoints, which does not record the trainer commit,
differs from the one saved with our run in these execution keys: \gls{AI2}'s
run used the non-fused loss, no flash attention, \texttt{torch.compile},
hybrid ZeRO-2 sharding with \num{32} replicas and block-and-size wrapping, and
two sequences per GPU, that is, \num{256} GPUs; our run used the fused loss,
flash attention, no compilation and \texttt{SHARD\_GRAD\_OP} sharding without
wrapping, as the published file specifies, and \num{32} sequences per GPU on
\num{16} GPUs. The remaining differences concern gradient accumulation,
in-loop evaluation, logging, checkpointing, resumption, file locations and run identity, plus two keys
that \gls{AI2}'s configuration lacks; apart from the data paths, no optimizer
hyperparameter and no scheduler, data, duration, loss-weight or precision
key differs. Part of the
shift of never-injected documents between the treated model and the control
(\Cref{sec:results-spillover}) could come from these execution differences
rather than from the injected data; that the largest shift falls on the
domain that received most of the injected tokens points to the data.

\paragraph*{Reference models}
The control is \gls{AI2}'s seed-42 stage-2 run, which \gls{AI2} labels
ingredient~3, read from the OLMo-native checkpoints under
\url{https://olmo-checkpoints.org/ai2-llm/peteish1/stage2/olmo2-1b-0425_stage2_ingredient3/}
every \num{1000} updates up to update \num{23000} and at update \num{23852}.
Their Hugging Face copies, the branches
\texttt{stage2-\allowbreak{}ingredient3-\allowbreak{}step}$N$\texttt{-tokens}$M$\texttt{B}
of
\texttt{allenai/\allowbreak{}OLMo-\allowbreak{}2-\allowbreak{}0425-\allowbreak{}1B},
agree bitwise with the downloaded files on every tensor we compared (four
tensors on each of five branches). All \num{24} configurations declare seed 42
and two sequences per GPU (\num{256} GPUs); steps \num{1000}--\num{19000} and
\num{20000}--\num{23852} carry the run names of two launches, whose
configurations differ only in run name, output folder and tracking name. The
seed-42069 run, used as the retrained reference (ingredient~1, branches
\texttt{stage2-\allowbreak{}ingredient1-\allowbreak{}step}$N$\texttt{-tokens}$M$\texttt{B}),
carries one run name on all \num{24} checkpoints and eight sequences per GPU
(\num{64} GPUs); its configuration differs from that of the seed-42 run only
in the seed, the per-GPU batch and gradient accumulation, the number of
sharding replicas, and its run name, output folders and tracking
names. The \texttt{main} revision of
\texttt{allenai/\allowbreak{}OLMo-\allowbreak{}2-\allowbreak{}0425-\allowbreak{}1B}
differs from the stage-1 endpoint and from the stage-2 endpoints of the
seed-42, seed-42069 and seed-666 runs, so every reference is loaded by branch
or checkpoint path.

\paragraph*{Post-training}
Post-training uses \gls{AI2}'s open-instruct at commit
\texttt{d19de7d2\allowbreak{}7ac2af2c\allowbreak{}d1dcbf85\allowbreak{}6dba1a2d\allowbreak{}077b37f4}
with its scripts \texttt{finetune.py},
\texttt{dpo\_\allowbreak{}tune\_\allowbreak{}cache.py} and
\texttt{grpo\_\allowbreak{}vllm\_\allowbreak{}thread\_\allowbreak{}ray\_\allowbreak{}gtrl.py}.
Our only code changes, in the first two scripts, keep the intermediate
checkpoints and replace the launch timestamp in the output-directory name with
a constant; the ZeRO-2 configuration file used for \gls{DPO}, which this
commit lacks, is a copy of \gls{AI2}'s
\texttt{configs/\allowbreak{}ds\_\allowbreak{}configs/\allowbreak{}stage2\_\allowbreak{}accelerate.conf}
at commit \texttt{fc7748e}. Our images are built from
\texttt{nvidia/\allowbreak{}cuda:12.4.1-\allowbreak{}cudnn-\allowbreak{}devel-\allowbreak{}ubuntu22.04}
with Python 3.10, PyTorch 2.6.0, transformers 4.50.0, DeepSpeed 0.15.4,
flash-attn 2.7.2.post1 and the other versions pinned in the commit's
\texttt{requirements.txt} except vLLM and xgrammar; the image for the
\gls{RLVR} stages adds vLLM 0.8.2, InfiniBand libraries and xgrammar 0.1.17
(the pinned 0.1.16 is no longer on PyPI). The Dockerfile of open-instruct at
this commit starts from a different base image and installs the same PyTorch
and flash-attn versions. The control starts from \gls{AI2}'s seed-42
step-\num{23852} checkpoint converted to Hugging Face format, whose weight
files are byte-identical to the branch
\texttt{stage2-\allowbreak{}ingredient3-\allowbreak{}step23852-\allowbreak{}tokens51B};
the treated model starts from our step-\num{23852} checkpoint, converted with
the same OLMo script. Both start models carry the released tokenizer files of
\texttt{allenai/\allowbreak{}OLMo-\allowbreak{}2-\allowbreak{}0425-\allowbreak{}1B}.
In the \gls{RLVR} stages, vLLM samples responses with temperature 1 and
top-$p$ 1; a response that does not end with the end-of-text token earns
reward 0 instead of its verifier score (GSM8K, MATH or instruction-following
constraints); and the Kullback--Leibler penalty uses the k3 estimator against
a frozen copy of the stage's starting model. Generation runs asynchronously
with the learner. Both \gls{RLVR} stages load the tokenizer of
\texttt{allenai/\allowbreak{}OLMo-\allowbreak{}2-\allowbreak{}1124-\allowbreak{}7B-\allowbreak{}DPO},
as \gls{AI2}'s commands do; its tokenizer files equal those of our \gls{SFT}
models. Prompts are drawn in an order shuffled with seed 1. For the treated
model and the control, the tokenized datasets are byte-identical at every
stage, and the configurations differ only in names, output directories,
starting model and tokenizer path.

Our \gls{SFT} averages the loss over the label tokens of each micro-batch of
two sequences (\texttt{reduce\_loss=mean}, the default at the pinned commit),
whereas \gls{AI2}'s logged OLMo~2~1B \gls{SFT} run (Weights~\&~Biases run
\texttt{ai2-llm/\allowbreak{}open\_\allowbreak{}instruct\_\allowbreak{}public/\allowbreak{}mihc7mab})
and its command in open-instruct's \texttt{docs/\allowbreak{}olmo2.md} at
commit \texttt{89b4f62} sum it (\texttt{reduce\_loss=sum}); the treated model
and the control both use the mean. Apart from this setting, the starting
model, paths and bookkeeping keys, our \gls{SFT} configuration equals that of
\gls{AI2}'s logged run, which started from a checkpoint that is not public.
Apart from the starting model, the dataset name, the tokenizer source, paths
and bookkeeping keys, our \gls{DPO} configuration equals that of \gls{AI2}'s
logged run (\texttt{k34ppe9q} in the same project), which also ran \num{2905}
steps; the logged dataset name now redirects to the dataset we used, and our
tokenizer, loaded from our \gls{SFT} model instead of
\texttt{allenai/\allowbreak{}OLMo-\allowbreak{}2-\allowbreak{}1124-\allowbreak{}13B},
has the \texttt{tokenizer.json} and \texttt{vocab.json} hashes that
\gls{AI2} logged. The \gls{RLVR} stages use \gls{AI2}'s original commands in
the same file flag for flag, apart from names, tracking, upload, Beaker
evaluation and local-path options. The logged configurations and code commit
of \gls{AI2}'s \gls{RLVR} runs are not public, and \gls{AI2}'s experiment
names state learning rates of \num{7e-7} and \num{9e-7} while its commands
pass \num{5e-7}, which we use. \gls{AI2} trained on H100 GPUs. Our \gls{RLVR}
stages ran on A100-SXM4-40GB GPUs, except the treated model's second stage,
which ran on A100-SXM4-80GB GPUs.

\paragraph*{Scoring}
The per-token log-probabilities on the evaluation set are computed with the
same token IDs, windows and positions for every model: OLMo-native checkpoints
are loaded in the mid-training image, and Hugging Face models in the
open-instruct image used for \gls{SFT} and \gls{DPO}. The intermediate
checkpoints of the treated run and of \gls{AI2}'s seed-42 and seed-42069 runs (updates \num{1000} to \num{23000},
\num{69} in all) were scored at document level only; all other models were
scored per token, and for them $\mu_t$ and $\sigma_t$ (\Cref{app:attacks})
are stored as well, computed over the full output layer of \num{100352} rows,
that is, the tokenizer's vocabulary and \num{74} padding
rows. The weights are cast to bf16 and the log-softmax is taken in fp32, with
one window per forward pass and no padding. The evaluation token IDs were
produced once with the Hugging Face tokenizer of
\texttt{allenai/\allowbreak{}OLMo-\allowbreak{}2-\allowbreak{}0425-\allowbreak{}1B},
without special tokens, and stored as int32; the IDs of the cohort-X model
samples and random tokens were generated directly. Every document ends with
one end-of-text token, and no beginning-of-sequence token is added. As the
first position of each window is not scored (\Cref{app:attacks}), a document
of $n$ tokens has $n-\lceil n/4096\rceil$ scored positions. Documents longer
than \num{65536} tokens are not scored; one evaluation document, a cohort-F
member of \num{69157} tokens, exceeds this limit and its pair is dropped. For
OLMo-native checkpoints, the attention kernel is taken from each checkpoint's
configuration: flash-attn for the treated mid-training checkpoints and
PyTorch's scaled-dot-product attention for \gls{AI2}'s checkpoints, including
the stage-1 endpoint. Every post-trained model is loaded in Hugging Face
format and scored with scaled-dot-product attention. Scoring draws no random
numbers; the treated and control endpoints were each scored twice on the
evaluation set as OLMo-native checkpoints, once without and once with $\mu_t$
and $\sigma_t$, and both passes give identical per-document scores. Each
scoring task on the evaluation set used one A100-SXM4-40GB GPU. The pre-treatment bits-per-byte
used for pairing were computed under the Hugging Face branch of the stage-1
endpoint with the same window conventions, but on text tokenized at scoring
time rather than on the stored token IDs.

\paragraph*{Analysis}
Arm assignment and the target positions of the copies are drawn from named
random streams under one seed (20260922). Resampling uses fixed seeds: the
data figures, the generated tables and the standardized ratio use seed
20261012 (\num{400} pair-bootstrap draws for \gls{AUC} and \gls{TPR},
\num{300} for the intervals in \Cref{fig:attacks}; \num{100000} sign vectors
for the randomization test; \num{20000} resamples per sample size for the
resampled power of the standardized ratio), intervals for mean effects use
\num{2000} pair resamples with seed 7, the trajectory windows \num{1000} with
seed 3, and the remaining analyses seed 20261013. The \gls{RLVR} rows of
\Cref{tab:stages}, the \gls{RLVR} values in the text and \Cref{tab:calibrated},
recomputed on all pairs with the evaluation functions used for the
standardized ratio, are computed separately. Apart from these, the data
figures, the generated tables and the standardized ratio were produced with Python 3.13.12,
NumPy 2.5.1, SciPy 1.17.1 and Matplotlib 3.10.8. A scan of the four post-training
datasets at the revisions in \Cref{tab:hyperparameters}, with the 8-word-shingle
matcher of the fingerprint search (\Cref{sec:design-verify}), flagging a
document at \num{40} or more shared contiguous words or \num{50} or more
shared distinct 8-grams, found \num{16} evaluation documents, all in
\gls{SFT} or \gls{DPO} examples; \num{11} of them belong to \num{10} pairs.
Both members of these pairs and the five flagged documents of the third,
never-injected arm (\Cref{sec:design-pairs}) are excluded from every reported
analysis.

\end{document}